\documentclass[preprint]{iucr}              

\usepackage{amsmath}
\usepackage{calc}
\usepackage{graphicx,color}
\usepackage{latexsym,amssymb}
\usepackage[mathscr]{eucal}
\usepackage{amsthm,amsxtra,amscd}
\usepackage{layout,bm,dcolumn}
\usepackage{cite,upref}

\newcommand{\Geo}[1]{\ensuremath{\mathcal{#1}}}

\newcommand{\DST}[1]{{\ensuremath{\displaystyle{#1}}}}
\newcommand{\DSF}[2]{{\ensuremath{\displaystyle{\frac{#1}{#2}}}}}

\newcommand{\CNJM}[1]{{\ensuremath{{\overline{#1}}}}}

\newcommand{\sinc}[1]{{\ensuremath{\mathrm{sinc}\left({#1}\right)}}}
\newcommand{\IMA}{{\ensuremath{\mathrm{i}}}}
\newcommand{\DD}[1]{{\ensuremath{\mathrm{d}{#1}}}}
\newcommand{\DDD}[3]{{\ensuremath{\mathrm{d}^{#1}{#2}}}}

\newcommand{\eref}[1]{Eq.~(\ref{#1})}

\newcommand{\eeeref}[3]{Eqs.~(\ref{#1},\ref{#2},\ref{#3})}

\newcommand{\sref}[1]{Sec.~\ref{#1}}
\newcommand{\cref}[1]{Chap.~\ref{#1}}
\newcommand{\tref}[1]{Tab.~\ref{#1}}
\newcommand{\fref}[1]{Fig.~\ref{#1}}

\newcommand{\vq}{\ensuremath{\bm{q}}}

\newcommand{\vuq}{\ensuremath{\bm{\hat{q}}}}
\newcommand{\vr}{\ensuremath{\bm{r}}}

\newcommand{\vy}{\ensuremath{\bm{y}}}
\newcommand{\vd}{\ensuremath{\bm{d}}}
\newcommand{\vud}{\ensuremath{\bm{\hat{d}}}}
\newcommand{\vuy}{\ensuremath{\bm{\hat{y}}}}
\newcommand{\lrb}[1]{\ensuremath{\left({#1}\right)}}
\newcommand{\lrs}[1]{\ensuremath{\left[{#1}\right]}}
\newcommand{\lrc}[1]{\ensuremath{\left\{{#1}\right\}}}
\newcommand{\lrv}[1]{\ensuremath{\left|{#1}\right|}}
\newcommand{\lrvv}[1]{\ensuremath{\left\|{#1}\right\|}}
\newcommand{\lra}[1]{\ensuremath{\left\langle{#1}\right\rangle}}
\newcommand{\EE}{\ensuremath{\mathrm{e}}}
\newcommand{\iq}{\ensuremath{{ {\mathcal{I}}\lrb{Q}}}}
\newcommand{\sq}{\ensuremath{{ {\mathcal{S}}\lrb{Q}}}}
\newcommand{\gr}{\ensuremath{{{\mathcal{G}}\lrb{r}}}}

\newcommand{\REVISED}[1]{\textcolor{black}{{#1}}}
\newcommand{\RE}{{\ensuremath{\mathfrak{Re}}}}
\newcommand{\IM}{{\ensuremath{\mathfrak{Im}}}}

\newtheoremstyle{note}
  {3pt}
  {3pt}
  {}
  {}
  {\itshape}
  {:}
  {.5em}
  {}
\theoremstyle{note}
     \paperprodcode{a000000}      
     \paperref{xx9999}            
     \papertype{FA}               
     \paperlang{english}          

     \journalcode{A}              
     \journalyr{2019}
     \journalreceived{\relax}
     \journalaccepted{\relax}
     \journalonline{\relax}

\begin{document}                  



\title{Texture corrections for total scattering functions}


\cauthor[a]{Antonio}{Cervellino}{antonio.cervellino@psi.ch}{address if different from \aff}
\author[b]{Ruggero}{Frison}

\aff[a]{Swiss Light Source, Paul Scherrer Institute, CH-5232 Villigen PSI, \country{Switzerland}}
\aff[b]{Center for X-ray Analytics, Empa Swiss Federal Laboratories for Materials Science and Technology, CH-8600 D\"ubendorf, \country{Switzerland}}









\maketitle                        

\begin{synopsis}
The Debye Scattering Equation (DSE) is generalised and augmented in order to account for moderate preferred-orientation (texture) effects 
which can be easily represented in terms of spherical harmonics. The modified DSE evaluates the differential cross section as a function of atomic coordinates and texture coefficients, subject to symmetry constraints. Implications on the evaluation of total scattering functions 
as the {\gr} (or Pair Distribution Function) as a direct transform of powder diffraction 
data from textured samples are also discussed. 
\end{synopsis}

\begin{abstract}
Many functional materials are today synthesised in form of nanoparticles displaying  preferred orientation effects to some small or large extent. The analysis of diffraction data of such kind of systems is best performed in the framework of the total scattering approach that prescinds from translation symmetry assumptions. We therefore derived modified expressions 
for the most common total scattering functions, in particular the Debye Scattering Equation (DSE) that 
yields the texture-averaged differential cross section as a function of atomic coordinates and texture parameters. The modified DSE encodes  higher-order even spherical Bessel functions which account for the texture effect. Selection rules arising from experimental geometries and symmetries are discussed. In addition the duality of the texture effect is introduced showing the effects of texture on both the {\iq} and {\gr}. The paper includes several definitions and appendices which are meant to be useful for those involved in the development of crystallographic computing.
\end{abstract}


\section{Introduction}\label{intro}

Preferred orientation (texture) is a complex effect that bridges powder diffraction to single crystal diffraction. In the last years in materials science there is an  increasing trend towards the synthesis and subsequent analysis of materials displaying only partial order often in the nanometer length scale in the form of nanoparticles (\citeasnoun{tekumalla_superior_2019}), thin-films (\citeasnoun{rijckaert_pair_2018}, \citeasnoun{Dippel_Local_2019}), or fiber-textured materials such as bone-like (\citeasnoun{tan_bio-inspired_2019}) or wood-like (\citeasnoun{lagerwall_cellulose_2014}). The analysis of these kinds of materials is best performed within the framework of the total scattering approach that prescinds from periodicity and therefore avoids Bragg formalism yet providing quantitative information on the structural parameters as well as on the size and shape of the scattering domain \citeasnoun{guagliardi_crc-concise_2015}. While intensity corrections for Bragg intensities are known (\citeasnoun{roe_description_1965}, \citeasnoun{bunge_texture_1969}, \citeasnoun{popa_texture_1992}), within the total scattering approach the problem of evaluating the {\sq} and {\gr} functions in the presence of texture has never been quantitatively tackled. {{Only some generalities have been presented in a preprint by \citeasnoun{GongBillinge2018}.}}
Therefore this paper deals with computation - \emph{via} an extended Debye scattering equation 
{\REVISED{(concisely DSE in this paper, see \citeasnoun{Debye1915})}} - of 1-D powder diffraction patterns obtained from crystalline powders having a non-uniform orientation distribution function (ODF hereafter).
We will remain in the realm of ``textured powders'' or powders with a weak to strong preferred orientation, but not so strong as to be better defined as mosaic crystal sets.
A complete treatment in the frameworks of spherical harmonics for the most common powder diffraction geometries is presented. The {\sq} can be computed by and extended version of the {\REVISED{DSE}} comprising now sums over spherical Bessel functions of all (even) orders. {\REVISED{Selection rules arising from symmetries 
and explicit expressions for the most common experimental geometries are given}}. 
Concerning the {\gr}, the effects of texture result in a fundamental indetermination that has important consequences, that we will discuss towards the end. 
We will start appropriately defining terms we use, although some of them may be familiar to part of the readers, and a brief recall of the part of the scattering theory relevant in this context (\sref{sec:scat_th}); at last in \sref{sec:odf} the definition of the analytic problem and its solution is given. Some useful mathematical functions (\sref{app}) and computing details (\sref{appB}) are reported in the Appendix.  

\subsection{Some definitions}\label{somedef}

\begin{itemize}\label{uno}
\item [1.\ ] {\emph{Atomic object} (AO) is a set of atoms rigidly bound together, constituting a particle, a nanoparticle, a molecule, a nanowire, a nanocrystal (NC), ... }
\item [2.\ ]{\emph{Powder} is an ensemble of a large number of identical AO constrained in a given volume and assuming all possible orientations in space with a certain probability density. In ideal powder the orientation probability distribution function is uniform and isotropic, 
but in reality there are many cases where it is not so.}
\item [3.\ ]{\emph{Symmetry} is the point (or Laue) symmetry group of the AO considered as a whole. This also (and especially) when the object is a (perhaps small) portion of a perfect crystal. 
In fact, translational symmetry cannot apply to a limited object. The crystal point group symmetry is the maximal possible symmetry group of the AO, even if the external shape would be more symmetric. 
So, a cubic cutout of a monoclinic crystal may be at most monoclinic, and that only if one of the cube axes coincides with the monoclinic 2-fold axis. }
\end{itemize}

\subsection{Additional considerations}

It must be also clear that, while the diffraction pattern of an ideal powder is essentially one-dimensional - because the 
intensity in reciprocal space varies only radially - a non-ideal powder (with) \emph{i.e.} with a non-uniform ODF (or \emph{textured} powder) has full 3-D dependence in reciprocal space. In fact, in an extreme case, all AO might be parallel and co-aligned - and if they were to be NC, the pattern would be essentially that of a single crystal. This might entail the need to measure more like a single crystal. 
We are interested mainly in the case where the uniformity of the ODF is only lightly perturbed. 
Then it makes still sense to measure the powder as such, with one of the traditional geometries (as discussed later). 
The variable in the patterns so measured is the deflection angle $2\theta$, or better the transferred momentum magnitude 
$q=2\sin\lrb{\theta}/\lambda$ (or $Q=4\pi\sin\lrb{\theta}/\lambda$), where $\lambda$ is the incident wavelength. As the differential scattering cross section for textured powders is not only a function of $q$ but in general of the vector $\vq$, the experimental geometry is essential in order to take into account texture effects. 
In particular, we must take into account additional symmetries arising from the sample nature and/or special averaging means applied. 
In many geometries, for instance, it is customary to rotate the sample around an axis while the measurement is taken; this will of course affect 
the texture, reducing it as a single axis rotation performs a partial orientation averaging. 
Therefore, we must specialize the concept of symmetry in two kinds:
\begin{itemize}
\item [A.\ ]{\emph{Object symmetry} that is the one defined in \sref{somedef} at point 3; }
\item [B.\ ]{\emph{Sample symmetry} that is the one defined just above.}
\end{itemize}
Clearly, the geometric relationship between the symmetry elements of the two kinds is important and needs to be specified.

\section{Scattering theory}\label{sec:scat_th}

Take an AO  composed of $N$ atoms indexed with $j=1\ldots N$, each centred at positions $\vr_j$ and each with 
isotropically variable spatial distributions $\beta_j\lrb{\lrv{ \vr-\vr_j}}$ of scattering length around $\vr_j$. Hereafter, 
a vector is denoted in bold and its length not (\emph{e.g.} $\lrv{\vr}=r$).
We assume hereafter that the coordinate system is chosen so that the main symmetry axes are along the coordinate axes. 
The scattering length density is then
\[
\rho\lrb{\vr}=\mathop{\sum}_{j=1}^N\beta_j\lrb{\lrv{\vr-\vr_j}}=\mathop{\sum}_{j=1}^N\int\DDD{3}{\vr'}
\,\delta\lrb{\vr'-\vr
+\vr_j}\beta_j\lrb{r'}
\]
Its Fourier transform is easily evaluated as
\[
F\lrb{\vq}=\int\DDD{3}{\vr}
\EE^{2\pi\IMA \vq\cdot\vr}\rho\lrb{\vr}=
\mathop{\sum}_{j=1}^N
\EE^{2\pi\IMA \vq\cdot\vr_j}
\int\DDD{3}{\vr'}\EE^{2\pi\IMA \vq\cdot\vr'}\beta_j\lrb{r'}
=
\mathop{\sum}_{j=1}^N
\EE^{2\pi\IMA \vq\cdot\vr_j}f_j(q)
\]
where we have set
\[
f_j(q)\equiv\int\DDD{3}{\vr'}\,\EE^{2\pi\IMA \vq\cdot\vr'}\beta_j\lrb{r'}
\]
The scattering factors $f_j$ are known and tabulated functions for all atoms and ions 
and for X-rays as well as for neutrons and electrons
(disregarding the weak perturbations due to the atomic environment). 
The common feature is that they are complex-valued but isotropic in reciprocal space. This is a consequence of the 
isotropy in direct space of the associated scattering length densities. 

The differential elastic coherent scattering cross section $I(\vq)$ is now proportional to 
the square modulus of $F\lrb{\vq}$, giving
\begin{eqnarray}
I\lrb{\vq}&=&\lrv{F\lrb{\vq}}^2=\lrv{
\mathop{\sum}_{j=1}^N
\EE^{2\pi\IMA \vq\cdot\vr_j}f_j(q)
}^2
\nonumber\\&=&\mathop{\sum}_{j,k=1}^Nf_j\CNJM{f}_k\EE^{2\pi\IMA \vq\cdot\lrb{\vr_j-\vr_k}}
\label{fullexp}\\&=&
\mathop{\sum}_{j=1}^N\lrv{f_j}^2\label{selfs}\\&+&
2\mathop{\sum}_{j>k=1}^N\RE\lrb{f_j\CNJM{f}_k}\cos\lrb{2\pi \vq\cdot\vd_{jk}}
\label{costerm}\\&-&
2\mathop{\sum}_{j>k=1}^N\IM\lrb{f_j\CNJM{f}_k}\sin\lrb{2\pi \vq\cdot\vd_{jk}}\label{anom}
\end{eqnarray}
Note that of the three resulting terms:
\begin{itemize}
\item [-- ] {the term in \eref{selfs}, the self-scattering, is isotropic (depends only on $q$) 
because (as it is often assumed) the atomic scattering length densities are so. Therefore this term does not change if the 
ODF is not uniform.}
\item [-- ] {the term in \eref{costerm}, let us name it \emph{principal scattering}, is even in $\vq$;} 
\item [-- ] {the term in \eref{anom}, for us \emph{secondary scattering}, is odd in $\vq$;} 
\end{itemize}
We shall be neglecting in the following the secondary scattering in expression (\ref{anom}). There are several reasons for that.
First, let us look at the magnitude of the scattering factors products. In the X-ray case, we have 
$f=f^{(0)}+f'+\IMA f''$, 
where $f^{(0)}$ (real) is the true elastic scattering term, depending only on $q$ 
and at small $q$ we have $f^{(0)}\sim Z$ (the atomic number); whilst 
$f'+\IMA f''$ constitute the \emph{anomalous} scattering factor part (real and imaginary parts), caused by the atomic electron binding, constant with respect to $q$ and varying only with the wavelength. See for reference and databases of atomic scattering factors \citeasnoun{XRDB2009}, 
\citeasnoun{EPDL97}, \citeasnoun{ATSCF5G}, \citeasnoun{NIST_web}, \citeasnoun{NIST1995}, \citeasnoun{NIST2000}. 
The structure factors are usually decomposed as
in standard conditions (far from elemental absorption edges) the ratios $\lrv{f'/f^{(0)}}$ and $\lrv{f''/f^{(0)}}$ are small. 
Then
\[
\RE\lrb{f_j\CNJM{f}_k}=(f^{(0)}_j+f'_j)(f^{(0)}_k+f'_k)+f''_jf''_k\approx f^{(0)}_jf^{(0)}_k;
\]\[
\IM\lrb{f_j\CNJM{f}_k}=(f^{(0)}_k+f'_k)f''_j-(f^{(0)}_j+f'_j)f''_k\approx f^{(0)}_kf''_j-f^{(0)}_jf''_k
\]
and it is clear that the imaginary part is small.
Only in special conditions the $f'$ or $f''$ can become large. 
Secondly, note that if the atomic species of the $j$-th and $k$-th atoms are the same clearly 
$\IM\lrb{f_j\CNJM{f}_k}=\IM\lrb{\lrv{f_j}^2}=0$, implying that the only contributing terms come from interatomic vectors linking atoms of different species. In monoatomic samples the secondary scattering will always be zero.
Third, consider the degree of preferred orientation. We range from ideal powder to single crystals, with many intermediates. 
Any even partial ODF averaging that mixes up $I\lrb{\vq}$ and $I\lrb{-\vq}$ will cancel partly or totally the secondary scattering. 
When the ODF is uniform (ideal powders), the odd sine terms average to exactly zero. This paper deals mainly with non-ideal powders, 
where the ODF is not uniform but also not as sharp as in a single crystal. For this reason, in most cases the secondary scattering can be neglected and one can assume that $I\lrb{\vq}$ is an even function of $\vq$. 
This has important consequences on the ODF averaging. 

We mention in passing that, as it must be, the effect in single crystals has been noted (Friedel pairs, \citeasnoun{Friedel1913}) and exploited for phasing, see \citeasnoun{Bijvoet1951}. 
{\REVISED{Moreover, this scattering contribution can be exploited in resonant conditions, that is, whenever the 
wavelength can be chosen so as to maximize it.}}

For completeness, we also give the inverse Fourier transform of \eref{fullexp} as it represents the pair correlation of the scattering 
density:
\begin{eqnarray}
\label{pacofusc1}
{\bm{\gamma}}\lrb{\vr}&=&\mathop{\sum}_{j,k=1}^N
\int\DDD{3}{\vr'}\,
\beta_j\lrb{\lrv{\vr'-\vd_{jk}}}\beta_k\lrb{\lrv{\vr+\vr'-\vd_{jk}}}
\\&=&\mathop{\sum}_{j,k=1}^N
\int\DDD{3}{\vr'}\,\int\DDD{3}{\vr''}\,
\delta\lrb{\vr''-\vr'+\vd_{jk}}
\beta_j\lrb{r''}\beta_k\lrb{\lrv{\vr+\vr''}}
\label{pacofusc2}
\end{eqnarray}

We now deal with evaluating the orientation average of terms like the sum in \eref{costerm}
when the ODF is not uniform (texture).

We shall follow the fundamental treatment of texture expressed in the basis of spherical harmonics, as in \citeasnoun{Roe65} and \citeasnoun{Bunge82}. For a detailed comparison of these two fundamental references see \citeasnoun{Esling82}. 
A new method has recently been presented (\citeasnoun{Mason08}; \citeasnoun{Mason09}), 
using the quaternion (axis-angle) {\REVISED{parametrization 
(\citeasnoun{QUAT-Moraw89}, \citeasnoun{QUAT-Kazan09}, \citeasnoun{QUAT-Karney07}, \citeasnoun{RADON_Bernstein05a}, 
\citeasnoun{RADON_Bernstein05})}} for 3D rotations instead of the less intuitive Euler matrices. 
We will not deal with this approach in this paper. 
{\REVISED{A very important paper for the treatment of symmetry is \citeasnoun{popa_texture_1992}, refined in \citeasnoun{popa_book_2008}; well known references are also \citeasnoun{Jarvinen_1993} and \citeasnoun{VonDreele_1997}. 
Special function definitions from \citeasnoun{NikifUva} and \citeasnoun{DLMF_web}}}

\section{Orientation Distribution Function formalism}\label{sec:odf}

An Orientation Distribution Function {\REVISED{(ODF)}}
\[
g(\phi_1,\Psi,\phi_2)
\]
{\REVISED{is function of three Euler angles. Let also}} 
\[
E(\phi_1,\Psi,\phi_2)
\]
the Euler matrix corresponding to a rotation of $\phi_1$ around the $z$ axis, followed by a rotation of $\Psi$ around the $y$ axis, followed by a rotation of $\phi_2$ around the new $z$ axis.
{\REVISED{An ODF}} is normalised to have unit average:
\begin{eqnarray}
1&=&\DSF{1}{8\pi^2}\int_0^{2\pi}\DD{\phi_1}
\int_0^{\pi}\sin\lrb{\Psi}\DD{\Psi}
\int_0^{2\pi}\DD{\phi_2}\ g(\phi_1,\Psi,\phi_2)
\end{eqnarray}
so that 
\(
\lrb{8\pi^2}^{-1} g(\phi_1,\Psi,\phi_2)
\)
is a probability density. 
The uniform isotropic case is when $g(\phi_1,\Psi,\phi_2)=1$. 
The ODF-weighted average of the principal scattering - the meaningful part of the differential cross section, as in \eref{costerm} - becomes
\begin{equation}
I_g(q) = 2\mathop{\sum}_{j>k=1}^N\RE\lrb{f_j\CNJM{f}_k}U\lrb{ \vq,\vd_{jk}}
\label{Istru}
\end{equation}
where
\begin{equation}
U\lrb{\vq,\vd}
\equiv
\DSF{1}{8\pi^2}\int_0^{2\pi}\DD{\phi_1}
\int_{-1}^{1}\DD{\cos\lrb{\Psi}}
\int_0^{2\pi}\DD{\phi_2}\ g(\phi_1,\Psi,\phi_2)
\cos\lrb{2\pi \vq\cdot E^{-1}(\phi_1,\Psi,\phi_2)\vd}
\end{equation}

\subsection{Uniform isotropic ODF case - the DSE}\label{sec:DSE}

In the uniform isotropic case ($g=1$), it is simple to verify that
\begin{equation}
U\lrb{\vq,\vd}=\DSF{\sin\lrb{2\pi qd}}{2\pi qd}=\sinc{2\pi qd}=j_0\lrb{2\pi qd}
\label{Esimp}
\end{equation}
where 
\begin{equation}
j_0(x)=\sqrt{\pi/2}\ \DSF{J_{1/2}(x)}{x^{1/2}}=\DSF{\sin\lrb{x}}{x}
\label{def_j0}
\end{equation}
is the spherical Bessel function of 0 order (for definitions an excellent online reference is \citeasnoun{DLMF_web}).
In this simplest - and fortunately very frequent - case, the expression of the orientation-averaged differential cross section of our AO
is simply
\begin{equation}
I\lrb{q}=
\mathop{\sum}_{j=1}^N\lrv{f_j}^2\ +\ 
2\mathop{\sum}_{j>k=1}^N\RE\lrb{f_j\CNJM{f}_k}j_0\lrb{2\pi q d_{jk}}
\label{eq:DSE}
\end{equation}
that is the {\REVISED{well-known DSE, first presented by \citeasnoun{Debye1915}}}.

\subsection{Arbitrary ODF case}\label{sec:TXT}

In more complex cases we have first to make one further simplification. If 
\begin{equation}
\vd=d\vud = d\lrb{\cos\lrb{\beta}\sin\lrb{\Phi},\sin\lrb{\beta}\sin\lrb{\Phi},\cos\lrb{\Phi}}
\label{def-d}
\end{equation}
and
\[
\vy=E^{-1}(\phi_1,\Psi,\phi_2)\vd=d\vuy = d\lrb{\cos\lrb{\gamma}\sin\lrb{\Xi},\sin\lrb{\gamma}\sin\lrb{\Xi},\cos\lrb{\Xi}}
\]
this still does not fully determine the Euler angles $\phi_1,\Psi,\phi_2$. In fact, a further rotation around $\vuy$ is possible. 
This does not affect anything, of course, therefore it is convenient to average it out. 
It is possible (see \citeasnoun{Roe65}, \citeasnoun{Bunge82}) to expand $g$ in Generalised Spherical Harmonics (GSH), whose definition we take from \citeasnoun{NikifUva}: 
\[
g\lrb{\phi_1,\Psi,\phi_2}=\mathop{\sum}_{l=0}^{+\infty}\mathop{\sum}_{m,n=-l}^{l}C_{l;m,n}\EE^{\IMA \lrb{m\phi_2+n\phi_1}}P_l^{mn}\lrb{\Psi}
\]
and we note conditions (\citeasnoun{Nadeau03}):
{{
\begin{equation}
C_{0;0,0}=1;\qquad \lrv{C_{l;m,n}}\leqslant 2l+1
\label{maxvalZ}
\end{equation}
(where the second inequality is just an upper bound, as tighter bounds are very difficult to compute in general)}} and then we execute the averaging of rotations around $\vuy$ (\emph{cf.} \citeasnoun{Roe65}, \citeasnoun{Bunge82}):
\[
\lra{g\Bigl.\Bigr.}_{\vud | \vuy} \lrb{\gamma,\Xi}= 
\mathop{\sum}_{l=0}^{+\infty}
\DSF{4\pi}{2l+1}
\mathop{\sum}_{m,n=-l}^{l}C_{l;m,n}
%
\lrb{-1}^{m+n}
{\overline{Y}}_l^{m}\lrb{\Phi,\beta}Y_l^{n}\lrb{\Xi,\gamma}
\]
where $Y_l^m$ are ordinary Spherical Harmonics (SPH). 
There are unfortunately many definitions used in various fields; 
the definition used here (and a comparison with other common definitions) is given in Appendix, see \sref{app2:LegeP}, \sref{app3:LegeF}, \sref{app4:SPH}. 
They are complex functions:
\[
Y_l^{n}\lrb{\theta,\phi}=X_l^m\lrb{\cos(\theta)}\EE^{\IMA m\phi}
\]
where for convenience we define
{\REVISED{
\begin{equation}
X_l^m\lrb{u}=\sqrt{\DSF{2l+1}{4\pi}\DSF{(l-m)!}{(l+m)!}}P_l^m\lrb{u}
\label{def:X}
\end{equation}
}}
where the associated Legendre functions {\REVISED{$P_l^m$}} are defined in \sref{app3:LegeF}.
It is also convenient to use the plane wave expansion in spherical harmonics 
\begin{eqnarray}
\cos\lrb{2\pi qd\,\vuy\cdot\vuq}&=&{4\pi}\mathop{\sum}_{p=0}^{+\infty}\lrb{-1}^p\ j_{2p}\lrb{2\pi qd}
\mathop{\sum}_{m=-2p}^{2p}
{\overline{Y}}_{2p}^m\lrb{\vuy}Y_{2p}^m\lrb{\vuq}\\
\sin\lrb{2\pi qd\,\vuy\cdot\vuq}&=&{4\pi}\mathop{\sum}_{p=0}^{+\infty}\lrb{-1}^p\ j_{2p+1}\lrb{2\pi qd}
\mathop{\sum}_{m=-2p-1}^{2p+1}
{\overline{Y}}_{2p+1}^m\lrb{\vuy}Y_{2p+1}^m\lrb{\vuq}
\end{eqnarray}
or (using the SPH addition theorem, see {\emph{e.g.} \citeasnoun{ArfkenAddTheo}):
{\REVISED{
\begin{eqnarray}
\cos\lrb{2\pi qd\,\vuy\cdot\vuq}&=&
\mathop{\sum}_{p=0}^{+\infty}\lrb{-1}^p{\lrb{4p+1}}\ j_{2p}\lrb{2\pi qd}P_{2p}\lrb{\vuy\cdot\vuq}\\
\sin\lrb{2\pi qd\,\vuy\cdot\vuq}&=&
\mathop{\sum}_{p=0}^{+\infty}\lrb{-1}^p{\lrb{4p+3}}\ j_{2p+1}\lrb{2\pi qd}P_{2p+1}\lrb{\vuy\cdot\vuq}
\end{eqnarray}
}}
where, if {\REVISED{we express $\vuq$ in polar coordinates}}
\[
\vuq=\lrb{\cos\lrb{\alpha}\sin\lrb{\Theta},\sin\lrb{\alpha}\sin\lrb{\Theta},\cos\lrb{\Theta}}
\]
{\REVISED{we can write}}
\[
\vuy\cdot\vuq=\cos\lrb{\alpha-\beta}\sin\lrb{\Xi}\sin\lrb{\Theta}+\cos\lrb{\Xi}\cos\lrb{\Theta}
\]
Now we integrate over $\Xi,\gamma$:
{\REVISED{
\begin{eqnarray}
U\lrb{\vq,\vd}&=&\DSF{1}{4\pi}\int_0^{2\pi}\DD{\gamma}\int_{-1}^1\DD{\lrb{\cos\lrb{\Xi}}}
\lra{g\Bigl.\Bigr.}_{\vud | \vuy}\lrb{\gamma,\Xi}\ 
\cos\lrb{2\pi qd\,\vuy\cdot\vuq}\\
&=&
\mathop{\sum}_{p=0}^{+\infty}
\DSF{4\pi}{4p+1}\lrb{-1}^p\ j_{2p}\lrb{2\pi qd}\quad\times\nonumber\\
&\times&
\mathop{\sum}_{m,n=-2p}^{2p}C_{2p;m,n}
%
\lrb{-1}^{m+n}
{\overline{Y}}_{2p}^{m}\lrb{\vud}
Y_{2p}^n\lrb{\vuq}
\end{eqnarray}
}}

\subsection{Symmetry constraints}

The theory of symmetry constraints on the complex coefficients $C_{l;m,n}$ has been developed in \citeasnoun{Bunge82}. {\REVISED{
A very clear and concise derivation  is found also in
\citeasnoun{popa_texture_1992} (with several small imprecisions), 
\citeasnoun{Jarvinen_1993}, \citeasnoun{VonDreele_1997}
and \citeasnoun{popa_book_2008}}}. In fact, we have already exploited the fact that $\cos\lrb{2\pi \vq\cdot\vd_{jk}}$ is real and centrosymmetric (\emph{i.e.} $\cos\lrb{2\pi \vq\cdot\vd_{jk}}=\cos\lrb{-2\pi \vq\cdot\vd_{jk}}$).
We can still exploit the fact that $g$ - and therefore also $\lra{g\Bigl.\Bigr.}_{\vud | \vuy}$ - is real.

Define thus the real spherical harmonics {\REVISED{$R_l^m$}}:
\begin{eqnarray}
m=0\ :\qquad R_l^0(\theta,\phi)&=&Y_l^0(\theta,\phi)=\sqrt{\DSF{2l+1}{4\pi}}P_l\lrb{\cos\lrb{\theta}}
;
\\
m>0,\  |m|\leqslant l\ :\qquad R_l^m(\theta,\phi)
&=&\DSF{1}{\sqrt{2}}\lrs{Y_l^m(\theta,\phi)+{\overline{Y}}_l^m(\theta,\phi)}
=\DSF{1}{\sqrt{2}}
\lrs{Y_l^m(\theta,\phi)+
(-1)^m
Y_l^{-m}(\theta,\phi)}\nonumber\\
&=&\sqrt{\DSF{2l+1}{8\pi}}
\lrb{\DSF{\lrb{l-m}!}{\lrb{l+m}!}}^{1/2}P_l^m\lrb{\cos\lrb{\theta}}
\cos\lrb{ m\phi}
;\nonumber\\
m<0,\  |m|\leqslant l\ :\qquad R_l^m(\theta,\phi)
&=&\DSF{-\IMA}{\sqrt{2}}\lrs{Y_l^m(\theta,\phi)-{\overline{Y}}_l^m(\theta,\phi)}
=\DSF{-\IMA}{\sqrt{2}}\lrs{Y_l^m(\theta,\phi)-
(-1)^m
Y_l^{-m}(\theta,\phi)}\nonumber\\
&=&\sqrt{\DSF{2l+1}{8\pi}}
\lrb{\DSF{\lrb{l-|m|}!}{\lrb{l+|m|}!}}^{1/2}P_l^{|m|}\lrb{\cos\lrb{\theta}}
\sin\lrb{ |m|\phi}\nonumber.
\end{eqnarray}
{\REVISED{In terms of the $X_l^m$ functions (\eref{def:X}) we have the compact forms
\begin{eqnarray}
m=0\ :\qquad R_l^0(\theta,\phi)
&=&X_l^0\lrb{\cos\lrb{\theta}};
\nonumber\\
m>0,\  |m|\leqslant l\ :\qquad R_l^m(\theta,\phi)
&=&\DSF{1}{\sqrt{2}}\ X_l^m\lrb{\cos\lrb{\theta}}
\cos\lrb{ m\phi}
;\\
m<0,\  |m|\leqslant l\ :\qquad R_l^m(\theta,\phi)
&=&\DSF{1}{\sqrt{2}}\ X_l^m\lrb{\cos\lrb{\theta}}\sin\lrb{ |m|\phi}.\nonumber
\end{eqnarray}
Now}} we can rewrite
\begin{eqnarray}
U\lrb{\vq,\vd}
&=&
\mathop{\sum}_{p=0}^{+\infty}
\DSF{4\pi}{4p+1}\lrb{-1}^p\ j_{2p}\lrb{2\pi qd}\times\nonumber\\&\times&
\mathop{\sum}_{m,n=-2p}^{2p}Z_{2p;m,n}
\lrb{-1}^{m+n}
R_{2p}^{m}\lrb{\vud}
R_{2p}^n\lrb{\vuq}
\label{thisisgen}
\end{eqnarray}
where $Z_{2p;m,n}$ are now real coefficients.

We will expand now on symmetry conditions as from \citeasnoun{Bunge82} and \citeasnoun{popa_texture_1992}.

\subsection{Sample symmetry}\label{samsym}


\begin{figure}\centering
 \includegraphics[width=0.58\textwidth]{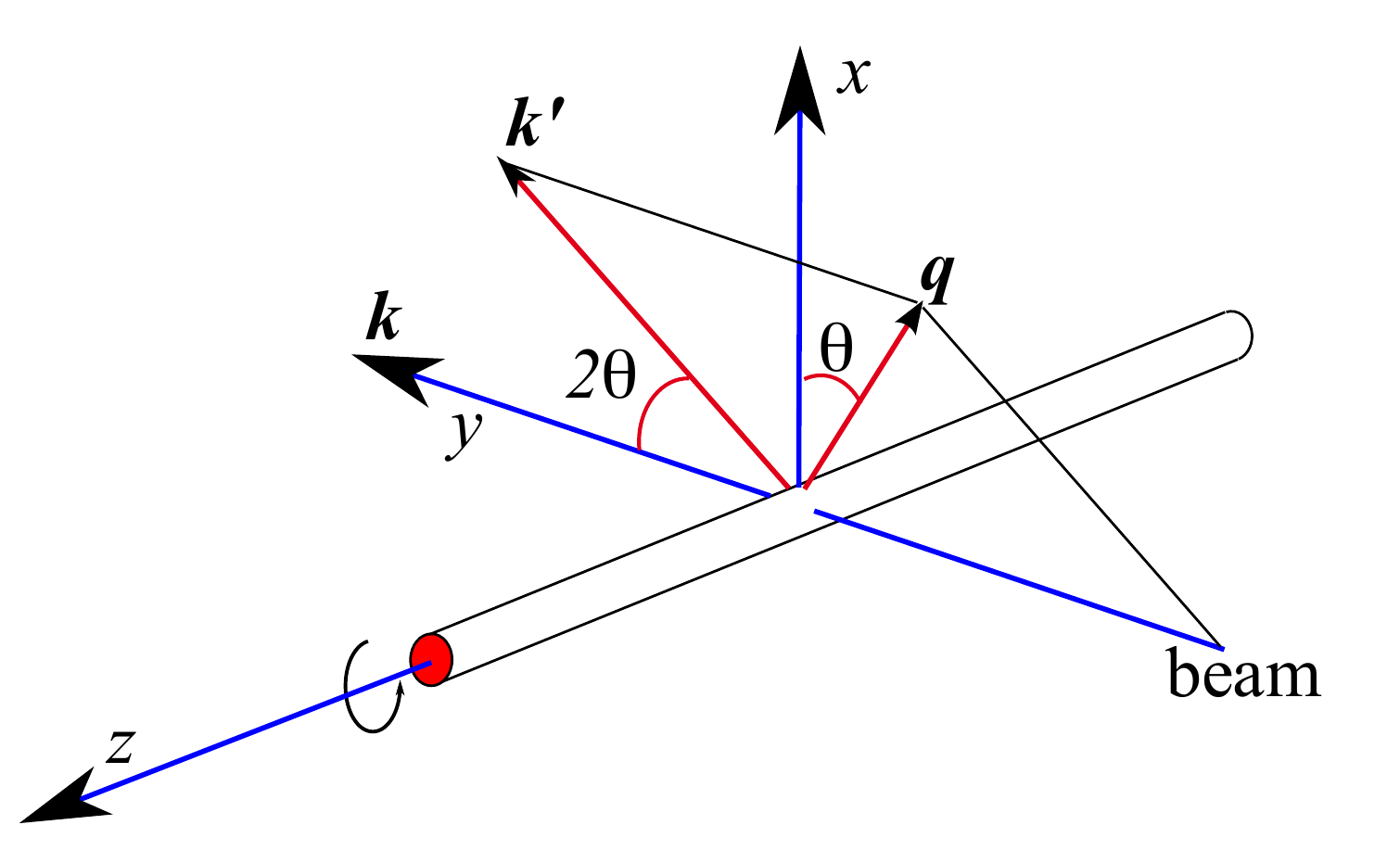}
 \includegraphics[width=0.58\textwidth]{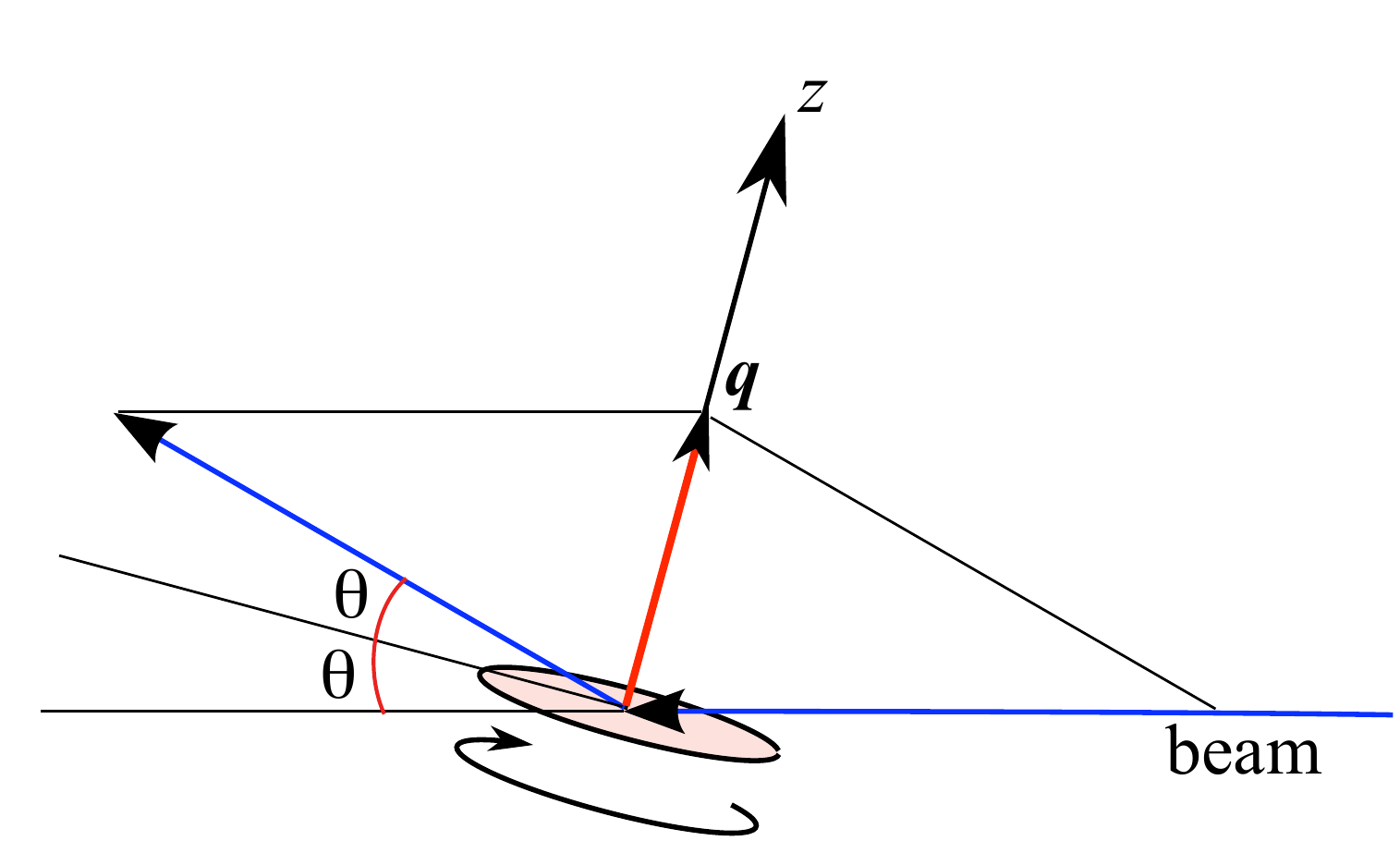}
 \includegraphics[width=0.58\textwidth]{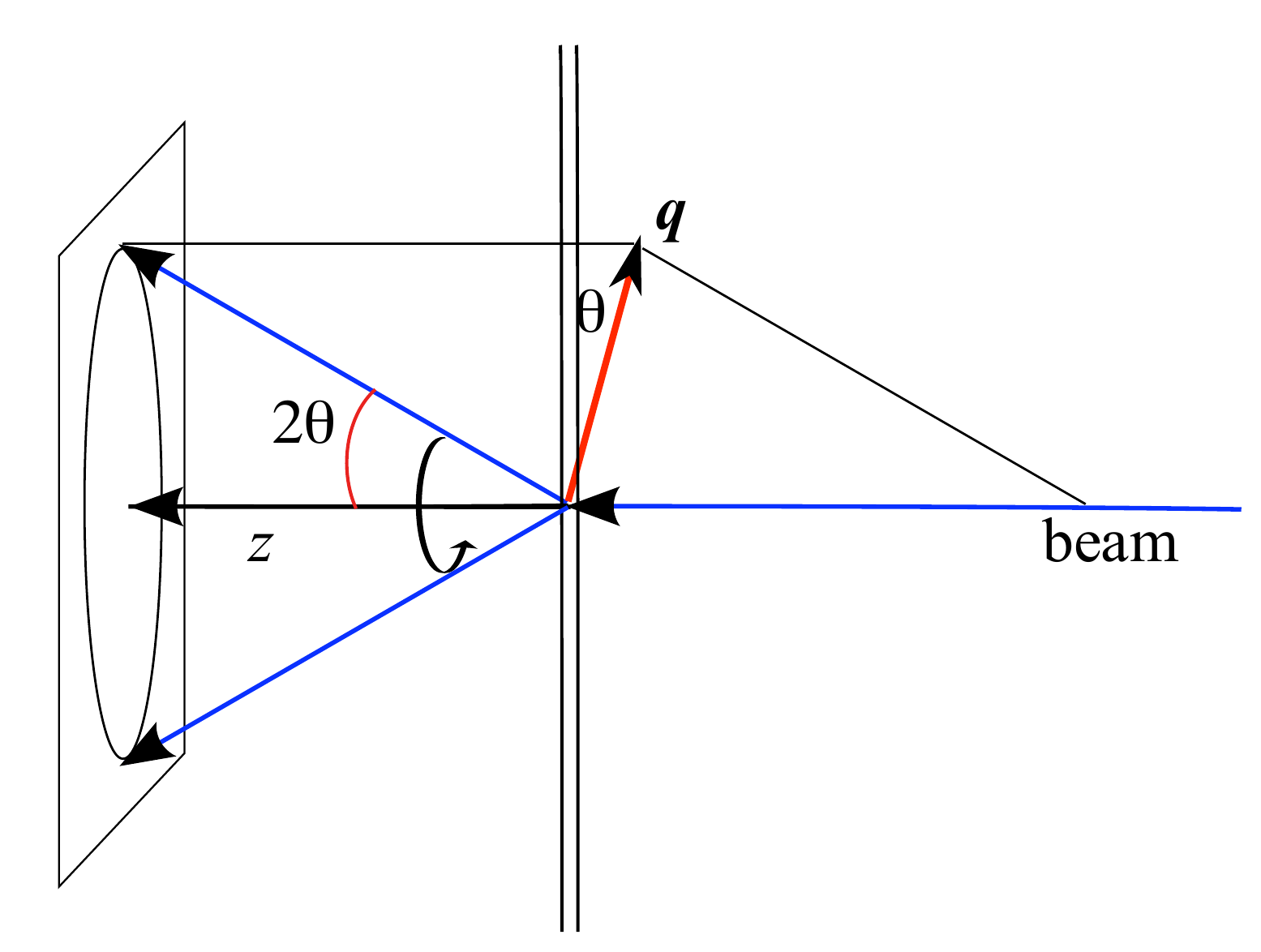}
\caption{(Top) Debye-Scherrer (capillary) geometry; (middle) Bragg-Brentano (symmetric reflection) geometry; (bottom) Flat-plate with frontal 2-D detector (transmission) geometry.}
\label{fig:geom}
\end{figure}

In the three most used experimental geometries for powder diffraction (DS or Debye-Scherrer with rotating capillary, BB or Bragg-Brentano with flat spinning plate, FP or flat-plate in transmission with frontal 2-D detector; see \fref{fig:geom}) we can assume cylindrical sample symmetry. In the first two cases this is due to the sample spinning around an axis which is then automatically the cylindrical symmetry axis; for FP, 
cylindrical symmetry ensues from integrating the Scherrer rings on the detector {{(or possibly, the flat plate could be made to rotate around the beam axis)}}.

We always set the $z$~axis along the cylinder axis. Then in the three cases, as it is evident from \fref{fig:geom}, 
\begin{eqnarray}
\text{DS:}\quad\vuq&=&\lrb{\cos\theta_d,-\sin\theta_d,0} \quad\rightarrow\quad\left\{
\begin{array}{rcl}
 \alpha & = &  -\theta_d \\
 \Theta & = &  \pi/2 
\end{array}
\right.
\\\nonumber
\text{BB:}\quad\vuq&=&\lrb{0,0,1} \quad\rightarrow\quad\left\{
\begin{array}{rcl}
 \alpha & = &  0 \\
 \Theta & = &  0
\end{array}
\right.
\\
\text{FP:}\quad\vuq&=&\lrb{\cos\theta_d,0,-\sin\theta_d} \quad\rightarrow\quad\left\{
\begin{array}{rcl}
 \alpha & = &  0 \\
 \Theta & = &  \pi/2+\theta_d
\end{array}
\right.
\end{eqnarray}

Then, as for cylindrical symmetry the only allowed value is always $n=0$,
\begin{eqnarray}
U\lrb{\vq,\vd}&=&
\mathop{\sum}_{p=0}^{+\infty}
\DSF{4\pi}{4p+1}\lrb{-1}^p\ j_{2p}\lrb{2\pi qd}
\mathop{\sum}_{m=-2p}^{2p}Z_{2p;m,0}
\lrb{-1}^{m}
R_{2p}^{m}\lrb{\vud}
R_{2p}^0\lrb{\vuq}
\\\nonumber&=&
\mathop{\sum}_{p=0}^{+\infty}
\sqrt{\DSF{4\pi}{4p+1}}\lrb{-1}^p\ j_{2p}\lrb{2\pi qd}
\mathop{\sum}_{m=-2p}^{2p}Z_{2p;m,0}
\lrb{-1}^{m}
R_{2p}^{m}\lrb{\vud}
P_{2p}\lrb{\cos\lrb{\Theta}}
\end{eqnarray}
Here we used the identity $P_{l}^0(x)=P_l(x)$. 
For the three geometries (DS, BB, FP) here considered,
\begin{itemize}
\item[Case DS:]{
\begin{eqnarray}
U\lrb{\vq,\vd}&=&
\mathop{\sum}_{p=0}^{+\infty}
\sqrt{\DSF{4\pi}{4p+1}}\lrb{-1}^p\ j_{2p}\lrb{2\pi qd}
\mathop{\sum}_{m=-2p}^{2p}Z_{2p;m,0}
\lrb{-1}^{m}
R_{2p}^{m}\lrb{\vud}
P_{2p}\lrb{0}
\nonumber\\&=&
\mathop{\sum}_{p=0}^{+\infty}
\sqrt{\DSF{4\pi}{4p+1}}\lrb{-1}^p\ j_{2p}\lrb{2\pi qd}
\mathop{\sum}_{m=-2p}^{2p}Z_{2p;m,0}
\lrb{-1}^{m}
R_{2p}^{m}\lrb{\vud}\times\nonumber\\
&\times&
\left\{
\begin{array}{rcl}
 p=0 & : &  1 \\
 p>0 & : &  \DST{\binom{2p-1}{p}\DSF{1}{(-1)^p2^{2p-1}}}
\end{array}
\right.
\end{eqnarray}
}
\item[Case BB:]{
\begin{eqnarray}
U\lrb{\vq,\vd}&=&
\mathop{\sum}_{p=0}^{+\infty}
\sqrt{\DSF{4\pi}{4p+1}}\lrb{-1}^p\ j_{2p}\lrb{2\pi qd}
\mathop{\sum}_{m=-2p}^{2p}Z_{2p;m,0}
\lrb{-1}^{m}
R_{2p}^{m}\lrb{\vud}
P_{2p}\lrb{1}
\nonumber\\
&=&
\mathop{\sum}_{p=0}^{+\infty}
\sqrt{\DSF{4\pi}{4p+1}}\lrb{-1}^p\ j_{2p}\lrb{2\pi qd}
\mathop{\sum}_{m=-2p}^{2p}Z_{2p;m,0}
\lrb{-1}^{m}
R_{2p}^{m}\lrb{\vud}
\end{eqnarray}
}
\item[Case FP:]{
\begin{eqnarray}
U\lrb{\vq,\vd}&=&
\mathop{\sum}_{p=0}^{+\infty}
\sqrt{\DSF{4\pi}{4p+1}}\lrb{-1}^p\ j_{2p}\lrb{2\pi qd}
\mathop{\sum}_{m=-2p}^{2p}Z_{2p;m,0}
\lrb{-1}^{m}
R_{2p}^{m}\lrb{\vud}
P_{2p}\lrb{\sin\theta_d}
\nonumber\\
&=&\mathop{\sum}_{p=0}^{+\infty}
\sqrt{\DSF{4\pi}{4p+1}}\lrb{-1}^p\ j_{2p}\lrb{2\pi qd}
\mathop{\sum}_{m=-2p}^{2p}Z_{2p;m,0}
\lrb{-1}^{m}
R_{2p}^{m}\lrb{\vud}
P_{2p}\lrb{{q\lambda}/{2}}\nonumber\\ &&
\end{eqnarray}
}
\end{itemize}
Here we used the even parity of the even Legendre polynomials $P_{2p}(x)=P_{2p}(-x)$, 
and Bragg's law $q=2\sin\theta_d/\lambda$. 

Given the obvious constraint $C_{0;0,0}=Z_{0;0,0}=1$, we can extract the $p=0$ term and simplify the rest. 
{{
We define another quantity for convenience:
\begin{equation}
\mathbb{Y}_{2p}\lrb{\vud}\equiv
\sqrt{\DSF{\pi}{4p+1}}
\mathop{\sum}_{m=-2p}^{2p}Z_{2p;m,0}\,
\lrb{-1}^{m}\,
R_{2p}^{m}\lrb{\vud}
\label{defq}
\end{equation}
Now we have, for DS:
\begin{eqnarray}
\label{DS1}
U\lrb{\vq,\vd}&=&j_{0}\lrb{2\pi qd}+
\mathop{\sum}_{p=1}^{+\infty}
\ j_{2p}\lrb{2\pi qd}\ 
\lrs{
\DST{\binom{2p-1}{p}\DSF{1}{2^{2(p-1)}}}
}
\mathbb{Y}_{2p}\lrb{\vud}
\end{eqnarray}
For BB:
\begin{eqnarray}\label{BB1}
U\lrb{\vq,\vd}&=&j_{0}\lrb{2\pi qd}+
\mathop{\sum}_{p=1}^{+\infty}
\ j_{2p}\lrb{2\pi qd}\ 
\lrs{
2(-1)^p}
\mathbb{Y}_{2p}\lrb{\vud}
\end{eqnarray}
For FP:
\begin{eqnarray}\label{FP1}
U\lrb{\vq,\vd}&=&j_{0}\lrb{2\pi qd}+
\mathop{\sum}_{p=1}^{+\infty}
\ j_{2p}\lrb{2\pi qd}
\lrs{
2(-1)^p\,P_{2p}\lrb{{q\lambda}/{2}}
}
\mathbb{Y}_{2p}\lrb{\vud}
\end{eqnarray}
}}

\subsection{Atomic Object symmetry}\label{AOsym}

If the atom cluster has additional symmetries, also the sum over $m$ can be reduced 
due to additional constraints (\citeasnoun{Bunge82}, \citeasnoun{popa_texture_1992}). 
Let us explore the comstraints for classical non-cubic crystal symmetries.

\subsubsection{One axis}\label{oneax}
With one symmetry axis only of order $r$ ($r=2,3,4,6$, for Laue groups $C_{2h}\equiv 2/m$, 
$C_{3i}\equiv \overline{3}$, $C_{4h}\equiv 4/m$, $C_{6h}\equiv 6/m$, respectively), supposedly oriented along $z$,  
we have that some of the $Z_{2p;m,0}$ ($p>0$) coefficients are zero. 
In particular, the surviving ones are
\[
Z_{2p;kr,0}, \qquad k\in\mathbb{Z};\quad -2p\leqslant kr\leqslant 2p
\]

\subsubsection{Two axes}}\label{twoax}
With one symmetry axis of order $r$ ($r=2,3,4,6$), supposedly oriented along $z$, 
and an additional 2-fold axis orthogonal to it, 
we have - additionally to the former condition -
that, for $p>0$, if $m$ even ($m=2s$), 
\[
Z_{2p;-|m|,0}=0
\]
and if $m$ odd ($m=2s-1$), 
\[
Z_{2p;|m|,0}=0
\]
\emph{i.e.} only the cosine terms survive when $m$ is even (respectively, the sine terms when $m$ is odd).
The results are summarised in \tref{tbl:1}.

\subsubsection{Three axes}\label{kuby}
This is the cubic case. Symmetrised harmonics for this case are not simply 
an appropriate subset of the real harmonics $R_l^m$; we must form appropriate linear combinations of them 
(with fixed $l$, of course). The original derivation is due to \citeasnoun{BethevdL47}. 
{{\eref{defq} will be changed into
\begin{equation}
\mathbb{Y}_{2p}\lrb{\vud}\equiv
\sqrt{\DSF{\pi}{4p+1}}
\mathop{\sum}_{\mu=1}^{H_p}Z_{2p;\mu,0}\,
K_{2p}^\mu\lrb{\vud}
\label{defqK}
\end{equation}
where $H_p$ is a (small) number of allowed terms for each $p$.}}
Denote these so-called kubic harmonics as $K_l^{j_l}\lrb{\theta,\phi}$, 
where $1\leqslant j_l\leqslant N_l$ is simply a counter.
Two Laue groups belong to the cubic case - $T_{h}\equiv m\overline{3}$ and 
$O_{h}\equiv m\overline{3}m$. For the first, conditions as for group $D_{2h}\equiv mmm$ hold; for the second, 
conditions as for group $D_{4h}\equiv 4/mmm$ hold. Additionally, for both, we must add the condition \citeasnoun{popa_texture_1992} 
\[
K_l^{j_l}\lrb{\theta,0} = K_l^{j_l}\lrb{0,\pi/2}
\]
For $l<4$ no terms are present. 
For $l=4$, for both groups $T_{h}\equiv m\overline{3}$ and 
$O_{h}\equiv m\overline{3}m$, we have one term (polar angle $\theta$, azimut $\phi$):
\[
K_4^1=\sqrt{\DSF{7}{12}}X_4^0\lrb{\cos\lrb{\theta}}+\sqrt{\DSF{5}{6}}X_4^4\lrb{\cos\lrb{\theta}}\cos\lrb{4\phi}
\]
{\REVISED{where the $X$ functions are defined in \eref{def:X}.}}
For $l=6$, for 
$O_{h}\equiv m\overline{3}m$, we have one term (polar angle $\theta$, azimut $\phi$):
\[
K_6^1=-\DSF{1}{2\sqrt{2}}
X_6^0\lrb{\cos\lrb{\theta}}+\sqrt{\DSF{7}{4}}X_6^4\lrb{\cos\lrb{\theta}}\cos\lrb{4\phi}
\]
and for group $T_{h}\equiv m\overline{3}$ there is additionally
\[
K_6^2=-\sqrt{\DSF{11}{8}}
X_6^2\lrb{\cos\lrb{\theta}}\cos\lrb{2\phi}+\sqrt{\DSF{5}{8}}X_6^6\lrb{\cos\lrb{\theta}}\cos\lrb{6\phi}
\]
For $l=8$, for 
$O_{h}\equiv m\overline{3}m$ as well as for group $T_{h}\equiv m\overline{3}$, we have just 
one term:
\[
K_8^1=\DSF{\sqrt{33}}{8}
X_8^0\lrb{\cos\lrb{\theta}}
+\sqrt{\DSF{7}{24}}X_8^4\lrb{\cos\lrb{\theta}}\cos\lrb{4\phi}
+\sqrt{\DSF{65}{96}}X_8^8\lrb{\cos\lrb{\theta}}\cos\lrb{8\phi}
\]
For $l=10$, for 
$O_{h}\equiv m\overline{3}m$, we have one term (polar angle $\theta$, azimut $\phi$):
\[
K_{10}^1=-8\sqrt{\DSF{6}{65}}
X_{10}^0\lrb{\cos\lrb{\theta}}
+\DSF{4}{\sqrt{11}}X_{10}^4\lrb{\cos\lrb{\theta}}\cos\lrb{4\phi}
+8\sqrt{\DSF{3}{187}}X_{10}^8\lrb{\cos\lrb{\theta}}\cos\lrb{8\phi}
\]
and for group $T_{h}\equiv m\overline{3}$ there is additionally
\[
K_{10}^2=-8\sqrt{\DSF{3}{247}}
X_{10}^2\lrb{\cos\lrb{\theta}}\cos\lrb{2\phi}
-8\sqrt{\DSF{6}{19}}X_{10}^6\lrb{\cos\lrb{\theta}}\cos\lrb{6\phi}
+8\sqrt{\DSF{2}{85}}X_{10}^{10}\lrb{\cos\lrb{\theta}}\cos\lrb{10\phi}
\]
For $l=12$, for 
$O_{h}\equiv m\overline{3}m$, we have two terms (polar angle $\theta$, azimut $\phi$):
\begin{eqnarray}
K_{12}^1&=&\DSF{20}{9}\sqrt{\DSF{41}{11}}
X_{12}^0\lrb{\cos\lrb{\theta}}
-\DSF{5}{2}\sqrt{\DSF{41}{91}}
X_{12}^4\lrb{\cos\lrb{\theta}}\cos\lrb{4\phi}
+10\sqrt{\DSF{82}{12597}}
X_{12}^8\lrb{\cos\lrb{\theta}}\cos\lrb{8\phi};
\nonumber\\
K_{12}^2&=&80\sqrt{\DSF{246}{676039}}
X_{12}^0\lrb{\cos\lrb{\theta}}
+80\sqrt{\DSF{82}{245157}}
X_{12}^4\lrb{\cos\lrb{\theta}}\cos\lrb{4\phi}
+80\sqrt{\DSF{41}{1771}}
X_{12}^8\lrb{\cos\lrb{\theta}}\cos\lrb{8\phi}+\nonumber\\
&&+\DSF{16}{5}\sqrt{\DSF{6}{41}}
X_{12}^{12}\lrb{\cos\lrb{\theta}}\cos\lrb{12\phi},\nonumber
\end{eqnarray}
and for group $T_{h}\equiv m\overline{3}$ there is additionally
\[
K_{12}^3=8\sqrt{\DSF{3}{17}}
X_{12}^2\lrb{\cos\lrb{\theta}}\cos\lrb{2\phi}
-\DSF{8}{5}\sqrt{\DSF{2}{7}}
X_{12}^6\lrb{\cos\lrb{\theta}}\cos\lrb{6\phi}
+8\sqrt{\DSF{6}{209}}
X_{12}^{10}\lrb{\cos\lrb{\theta}}\cos\lrb{10\phi}
\]
All kubic harmonics above are orthonormal.\\[5mm]
\begin{table}
\begin{center}
\label{tbl:1}
    \begin{tabular}{|lll|ccccccccccccc|}
        \hline\hline
        \multicolumn{3}{|r}{$2p=l\geqslant \lrv{m}=$}&0&1&2&3&4&5&6&7&8&9&10&11&12\\        \hline
$C_i$ & $\overline{1}$ &triclinic &0 &$\pm$&$\pm$&$\pm$&$\pm$&$\pm$&$\pm$&$\pm$&$\pm$&$\pm$&$\pm$&$\pm$&$\pm$\\
$C_{2h}$ & $2/m$ &monoclinic &0& &$\pm$& &$\pm$& &$\pm$& &$\pm$& &$\pm$& &$\pm$\\
$D_{2h}$ & $mmm$ &orthorhombic &0& &$+$& &$+$& &$+$& &$+$& &$+$& &$+$\\
$C_{3i}$ & $\overline{3}$ &trigonal 1 &0& & &$\pm$& & &$\pm$& & &$\pm$& & &$\pm$\\
$D_{3d}$ & $\overline{3}m$ &trigonal 2 &0& & &$-$& & &$+$& & &$-$& & &$+$\\
$C_{4h}$ & $4/m$ &tetragonal 1 &0& & & &$\pm$& & & &$\pm$& & & &$\pm$\\
$D_{4h}$ & $4/mmm$ &tetragonal 2 &0& & & &$+$& & & &$+$& & & &$+$\\
$C_{6h}$ & $6/m$ &hexagonal 1 &0& & & & & &$\pm$& & & & & &$\pm$\\
$D_{6h}$ & $6/mmm$ &hexagonal 2 &0& & & & & &$+$& & & & & &$+$\\
        \hline
$C_{\infty h}$ & $\infty/m$ &cylindric &0& & & & & & & & & & & & \\
        \hline
        \multicolumn{3}{|r}{$2p=l=$}& &1&2&3&4&5&6&7&8&9&10&11&12\\
        \hline
        \ & &&&&&&&&&&&&&&\ \\
$T_{h}$ & $m\overline{3}$ &cubic I & &&&&$K_4^1$ &&$K_6^1$ &&$K_8^1$&&$K_{10}^1$ &&$K_{12}^1$ \\
\multicolumn{3}{|l|}{(tetrahedral)} &&&&& &&&&&&$K_{10}^2$ &&$K_{12}^2$ \\
& & &&&&& &&&&&&&&$K_{12}^3$ \\
        \ & &&&&&&&&&&&&&&\ \\
$O_{h}$ & $m\overline{3}m$ &cubic II & &&&&$K_4^1$ &&$K_6^1$ &&$K_8^1$&&$K_{10}^1$ &&$K_{12}^1$ \\
 \multicolumn{3}{|l|}{(octahedral)}&&&&&&&&&&&&&$K_{12}^2$ \\
        \ & &&&&&&&&&&&&&&\ \\
\hline
    \end{tabular}
\end{center}
\end{table}

\section{Computation}\label{sec:comp}

The computation of the classical Debye scattering equation is made much easier by using the Gaussian sampling method, see \citeasnoun{NOI06jcc}, 
\citeasnoun{guagliardi_crc-concise_2015}. 
We briefly recall its principle. 
Firstly, we assume that either the system is monoatomic, or the sum over atom pairs in \eref{Istru} has been split in parts corresponding to each pair of atomic species. 
In this way we can factor out the possibly $q$-dependent scattering lengths products $\RE\lrb{f_j\CNJM{f}_k}$, that would then be multiplied after evaluating the partial sums over different pairs, to be finally summed at the very end. 
So in this part we will omit the scattering lengths products.

Given an interatomic distance $\vd_{jk}$, 
its contribution to the powder pattern is $j_0\lrb{2\pi qd_{jk}}$. 
As the $d_{jk}=\lrv{\vd_{jk}}$ values are a huge number (the square of the number of atoms in the AO) 
and they are all concentrated in a finite interval $0<d_{jk}< D$, with $D$ the diameter 
(or maximal linear dimension) of the AO, it pays off to consider only a discrete and uniformly spaced set of distances $d_m=m\Delta$ with 
appropriately chosen weights $\mathbb{W}_m$ and then compute the pattern as 
\begin{equation}\label{Sform}
C(q)\mathop{\sum}_{m=1}^{M_{max}}\mathbb{W}_m\,\, j_0\lrb{2\pi qm\Delta}
\end{equation}
instead of a much larger sum over terms like in \eref{eq:DSE}. 
Recalling briefly the procedure, each term is replaced by 
\begin{eqnarray}
&\ &C(q)
\DSF{\Delta}{\rho\sqrt{2\pi}d_{jk}}
\mathop{\sum}_{m=\max\lrb{0,\lfloor d_{jk}/\Delta\rceil-\zeta}}^{\lfloor d_{jk}/\Delta\rceil+\zeta}
\,j_0\lrb{2\pi qm\Delta}\times\nonumber\\&\times&
\lrs{
\exp\lrb{-\DSF{\lrb{m\Delta-d_{jk}}^2}{2\rho^2\Delta^2}}
-\exp\lrb{-\DSF{\lrb{m\Delta+d_{jk}}^2}{2\rho^2\Delta^2}}
}
\end{eqnarray}
Here, $\lfloor x\rceil$ is the nearest integer to $x$; $C(q)=\exp\lrb{2\pi^2\rho^2\Delta^2q^2}$ is a correction factor; $\rho=2.701$ is a numerical constant; 
$\zeta$ is an integer (typically $\zeta=30 \ \text{to}\ 60$) such that (numerically) $\exp\lrb{-\lrb{m\Delta+d_{jk}}^2/\lrb{2\rho^2\Delta^2}}$ can be considered negligibly small.  The second Gaussian centred at $-d_{jk}$ is almost always negligible except when $m$ is close to 0. Finally, the parameter $\Delta$ - the sampling step - must be chosen so that $\Delta<1/(2q_{max})$, where $q_{max}$ is the maximum momentum transfer value in the pattern to be calculated; a numerically safe choice is $\Delta\leqslant 0.4/q_{max}$. The Whittaker-Nyquist-Kotelnikov-Shannon upper limit for the sampling step (\citeasnoun{Shannon1949}) is also $1/(2q_{max})$, see \citeasnoun{guagliardi_crc-concise_2015} and \citeasnoun{Spring16}. Hence, 
this is the most efficient approximate method with negligible error (practically zero). 
Values of $\Delta$ ranging from 1~\AA{} to 0.03~\AA{} cover most imaginable powder diffraction experimental conditions with neutrons and X-rays. For an exhaustive derivation see \citeasnoun{NOI06jcc}. 
When adding more contributions to the pattern, the $q$-dependent factor $C(q)$ can be omitted and be left to be multiplied at the end. 
The contributions from different distances can be summed on the $\lrc{
m\Delta,\ m=1,\ldots,M_{max}+\zeta}$ grid and the pattern is built by accumulation, resulting in the 
$\mathbb{W}_m$ that multiply the $j_0\lrb{2\pi m\Delta q}$ contributions in \eref{Sform}. 

It is clear that such computational advantage can be preserved in the extended form. 
We now will explain how the procedure must be modified. 

We rewrite here the sum \eref{Istru} (assuming \eref{Esimp}) in a more convenient way
\begin{equation}
\mathop{\sum}_{j\neq k=1}^N j_0\lrb{2\pi q d_{jk}}=\mathop{\sum}_{\ell=1}^{\overline{N}_d}\mathcal{W}_\ell\ 
j_0\lrb{2\pi q d_\ell}
\label{eq:DSEw}
\end{equation}
where the set of interatomic vectors $\lrc{\vd_{jk}\,|\,j\neq k}$ has been split into ${\overline{N}_d}$ equivalence classes 
of interatomic vectors having the same length, each $\ell$-th class defined as $\lrc{\vd_{jk}\,|\,j\neq k; \ d_{jk}=d_\ell}$, $\ell=1,\ldots,{\overline{N}_d}$. 
The $j_0$ terms to be computed (or sampled) are only those containing the $d_\ell$ in argument. 
Each of them is weighted by $\mathcal{W}_\ell$, each being the number of $\vd_{jk}$ within the $\ell$-th equivalence class.

When having to compute superior orders, like in \eref{DS1}, \eref{BB1} or \eref{FP1}, or the functions defined in \sref{kuby}, 
the same equivalence classes define the $j_{2p}$ terms to be computed. 
Only the corresponding weights become more complex. In fact, now they depend also on the direction vectors $\vud_\ell$ belonging to the corresponding class, through the real SPH $R^m_{2p}\lrb{\vud_\ell}$. There one needs just to follow the relevant equation. We give next, however, some indication on how to compute economically the angular dependent terms.


\subsection{Angular functions computation}

Take a distance vector $\vd_\ell$ belonging to one of the equivalence classes defined above. 
Let
\[
\vd=d\lrb{\cos\lrb{\beta}\sin\lrb{\Phi},\sin\lrb{\beta}\sin\lrb{\Phi},\cos\lrb{\Phi}}=\lrb{x,y,z}
\]
(\emph{cf.} \eref{def-d}) given both in polar and in Cartesian coordinates with respect to the appropriate reference system. 
Defining for convenience
\[
\bm{\rho}\equiv\lrb{x,y,0},\qquad \rho=\sqrt{x^2+y^2}
\]
we can write the following interrelations:
\begin{eqnarray}
d&=&\sqrt{x^2+y^2+z^2}\\
\cos\lrb{\Phi} & = & \DSF{z}{d}; \qquad \sin\lrb{\Phi}=\sqrt{1-\cos^2\lrb{\Phi}}\\
\cos\lrb{\beta}&=&\DSF{x}{\rho};\qquad \sin\lrb{\beta}=\DSF{y}{\rho}
\end{eqnarray}
These are the only necessary relationships. 
For completeness we give the expressions for the angular values of $\Phi,\beta$, even if they are
not necessary:
\[
\Phi=\arccos\lrb{\DSF{z}{d}}
\]
\[
\beta =  \arctan\lrb{y,x}=
\left\{
\begin{array}{lcl}
\arctan\lrb{\DSF{y}{x}}-\pi\DSF{ \lrb{\text{sign}(x)-1}}{2} & \text{if}  &  x\neq 0 \\
  & \  &\   \\
 \DSF{\pi}{2}\text{sign}(y) & \text{if}  &  x= 0
\end{array}\right.
\]
where of course $\text{sign}(x)=x/\lrv{x}$. 

The direct values of the angles are not necessary because  in \eref{thisisgen}  the spherical harmonics depend only on sines and cosines of  
 $\Phi,\beta$ and of their integer multiples. 
 The latter can be most conveniently computed by using the relations
 \begin{eqnarray}
\cos\lrb{n\varphi}&=&T_n\lrb{\cos\lrb{\varphi}};\\
\sin\lrb{n\varphi}&=&\sin\lrb{\varphi}U_{n-1}\lrb{\cos\lrb{\varphi}}\nonumber
\end{eqnarray}
involving the Chebyshev polynomials of the first kind $T_n(x)$ (\citeasnoun{WolframTcheb}) and those of the second kind $U_n(x)$ 
(\citeasnoun{WolframUcheb}).  
{\REVISED{These are very conveniently and efficiently evaluated by recurrence relations. This is detailed in 
\sref{app:Clen}. Moreover, \sref{app:step2} deals with the case - frequent in this context - 
where only odd or even terms must be used.
So all computations can be performed without using any direct or inverse trigonometric functions. This 
enhancement is used to great effect in the DEBUSSY software suite (\citeasnoun{Debussy2p0_2015}), 
greatly speeding up computations of the DSE.}}

\section{Direct space direct transforms}\label{sec:pdf}

By means of a specialised Fourier transform of a powder diffraction pattern, it is possible to obtain 
a pattern in direct space, with a single radial coordinate $r$, showing a sharp peak wherever there are interatomic distances equal 
to $r$, whose height is related to the multiplicity of the distance and the scattering length product of the connected atoms. 
This is the basis of the well known PDF method {\REVISED{(\citeasnoun{ZernPrins1927}, \citeasnoun{Egami03}, \citeasnoun{billinge_book_2008})}} . The radial pattern in direct space is generally referred to as PDF (Pair Distribution Function), 
meaning that, in the sense roughly sketched above, it provides a weighted representation of the pair distances between atoms. 

While different functions are commonly used for the direct space representation, the most frequently associated with the 
PDF acronym are, since \citeasnoun{ZernPrins1927}, 
\begin{equation}
{\mathcal{G}}_t \lrb{r} = 
\lrs{4\pi r^{t-1} n_0}
\DSF{1}{2\pi^2}\int_0^{+\infty}Q\,\DD{Q}\sin\lrb{Qr}\lrb{\Geo{S}\lrb{Q}-1}, \qquad t=0,1, 2\ 
\end{equation}
with the more usual choice of variable $Q=2\pi q$. 
Most usually with the choice $t=1$  that we will assume in the following (${\mathcal{G}} \lrb{r}\equiv {\mathcal{G}}_1 \lrb{r}$); 
and here $n_0\, {=N/V}$ is the point density of atoms per unit volume, {{$V$ being the volume occupied by the AO}}. 
The $S(Q)$ function appearing there is just
\[
\Geo{S}\lrb{Q} = \DSF{I(Q)}{N\lra{\lrv{f}^2}},\quad\text{where}\quad
\lra{\lrv{f}^2}\equiv  \DSF{1}{N}\mathop{\sum}_{j=1}^N\lrv{f_j}^2
\]
where $I(Q)$ is the isotropic averaged differential cross section (\eref{eq:DSE}) expressed in the variable $Q$. More in detail,
\begin{eqnarray}
\Geo{S}\lrb{Q} &=& \DSF{1}{N\lra{\lrv{f}^2}}\mathop{\sum}_{j=1}^N\lrv{f_j}^2\ +\ 
\DSF{2}{N\lra{\lrv{f}^2}}\mathop{\sum}_{j>k=1}^N\RE\lrb{f_j\CNJM{f}_k}j_0\lrb{Q d_{jk}}\nonumber\\
&=&1+
\DSF{1}{N\lra{\lrv{f}^2}}\mathop{\sum}_{j\neq k=1}^N\RE\lrb{f_j\CNJM{f}_k}j_0\lrb{Q d_{jk}}
\end{eqnarray}
so
\begin{eqnarray}
{\mathcal{G}} \lrb{r} &=& 
n_0
\DSF{2}{\pi}\int_0^{+\infty}Q\,\DD{Q}\sin\lrb{Qr}
\mathop{\sum}_{j\neq k=1}^N\DSF{\RE\lrb{f_j\CNJM{f}_k}}
{N\lra{\lrv{f}^2}}
j_0\lrb{Q d_{jk}}\\&=&
\DSF{2r}{\pi V}
\mathop{\sum}_{j\neq k=1}^N
\int_0^{+\infty}Q^2\,\DD{Q}\,
\lrs{\DSF{\RE\lrb{f_j\CNJM{f}_k}}
{\lra{\lrv{f}^2}}}
 j_0\lrb{Qr}
j_0\lrb{Q d_{jk}}
\end{eqnarray}
In the simplest case where the term in square bracket is independent of $Q$, we can extract it from the integral and
\begin{eqnarray}
{\mathcal{G}} \lrb{r} &=&
\DSF{2r}{\pi V}
\mathop{\sum}_{j\neq k=1}^N
\lrs{\DSF{\RE\lrb{f_j\CNJM{f}_k}}
{\lra{\lrv{f}^2}}}
\int_0^{+\infty}Q^2\,\DD{Q}\,
 j_0\lrb{Qr}
j_0\lrb{Q d_{jk}}\nonumber\\
&=&
\DSF{1}{V}
\mathop{\sum}_{j\neq k=1}^N
\lrs{\DSF{\RE\lrb{f_j\CNJM{f}_k}}
{\lra{\lrv{f}^2}}}
\DSF{
\delta\lrb{r-d_{jk}}
}{d_{jk}}
\end{eqnarray}
Here we use the integral from \citeasnoun{DLMF_web}, Eq.~1.17.14: 
\[
\int_0^{+\infty}s^2\,\DD{s}\,j_l(xs)\,j_l(x's)=\DSF{\pi}{2xx'}\delta\lrb{x-x'}, \qquad l=0,1,2,\ldots
\]
One interesting side note is that, if we define a scalar product between complex functions on $\mathbb{R}^+=(0,+\infty)$:
\begin{equation}
\lra{f\Bigl|\bigr. g} \equiv
\int_0^{+\infty}4\pi s^2\,\DD{s}\, {\overline{f}}(s)g(s)
\end{equation}
it is immediate that this induces a norm
\[
\lrvv{f}=\lra{f\Bigl|\bigr. f}^{1/2}
\]
and a distance
\[
D\lrs{f,g}=\lrvv{f-g}
\]
So if we take the space $\mathbb{U}$ of well-behaved complex functions on $\mathbb{R}^+$, 
for instance those having finite norm and $C^\infty$ on $\mathbb{R}^+$, we can define its closure, 
the space of complex functionals on $\mathbb{U}$ as a Riesz space. 
Now, we write a slightly modified integral
\begin{equation}\label{prod0}
\lra{j_0\lrb{2\pi sr}\Bigl|\Bigr.
j_0\lrb{2\pi sr'}
}=\int_0^{+\infty}4\pi s^2\,\DD{s}\,j_0(2\pi sr)\,j_0(2\pi sr')=
\DSF{\delta\lrb{r-r'}}{4\pi rr'}
\end{equation}
This means that the functions
\[
j_0(Qr)=\DSF{\sin\lrb{Qr}}{Qr}
\]
constitute a complete orthogonal system on $\mathbb{R}^+$. In particular, 
the superior orders $j_l(Qr)$, $l>0$, can be expressed as linear combinations of the $j_0(Qr)$. 
Therefore there arises ambiguity in evaluating the {\gr} for a system with texture, because the higher orders of spherical Bessel functions 
will mix up in the {\gr} evaluated from experimental data.
A {\gr} curve from a textured powder will have to be carefully compared to an atomic model including the texture parameters, 
and even so the results may be ambiguous. 

\subsection{{\gr} from higher-order even spherical Bessel functions}\label{hiord}

As a last point, as it is not easy to find them in the literature, we give here expressions of the scalar product of the $j_0(Qr)$ basis functions with the 
$j_{2p}\lrb{Qr}$ even higher-order spherical Bessel functions that appear in the texture-generalised DSE. 
The only reference we could find is a paper by \citeasnoun{Maximon1991}. From there, with a bit of bookkeeping, 
\begin{eqnarray}
\lra{j_0\lrb{2\pi qr}\Bigl|\Bigr.
j_{2p}\lrb{2\pi qd}
}&=&\int_0^{+\infty}4\pi q^2\,\DD{q}\,j_0(2\pi qr)\,j_{2p}(2\pi qd)\nonumber\\
&=&
\DSF{(-1)^p}{4\pi rd}
\lrc{
\delta\lrb{r-d}
-\DSF{1}{d}\lrs{
\DSF{\DD{}}{\DD{x}}P_{2p}(x)
}_{x=r/d}\Theta\lrb{d-r}
}
\end{eqnarray}
The first term with the Dirac delta, apart from the sign $(-1)^p$, is identical to the result for $p=0$, see \eref{prod0}. 
This term is creating an infinitely sharp peak at $r=d$. Real-world samples show in fact sharp peaks, although not infinitely sharp because of 
atomic form factors, uncorrelated thermal vibrations and possibly disorder. Note, however, that the Dirac delta terms encoded in the higher-order spherical Bessel functions $j_{2p}$ have all the same intrinsic magnitude $(4\pi rd)^{-1}$, 
but alternating signs $(-1)^p$. Therefore, the interatomic distance peaks of the {\gr} will change in height due to texture, as a first-order effect; and it is very well possible that, for some combination of texture coefficients, some distance peaks might disappear. This is the dual of a 
similar well-known effect on the reciprocal space pattern - texture modifies the Bragg peak intensities and in 
some cases cancellation of some families of peaks has been observed. See \fref{fig:pdf} for some example calculated {\gr}.

Another effect comes from the 
second term, that has as factor a Heaviside function 
\[
\Theta(x)=\left\{
\begin{array}{ccc}
  1 & \text{if}  & x>0   \\
 1/2 & \text{if}  & x=0 \\
  0 & \text{if}  & x<0     
\end{array}
\right.
\]
that reduces to 0 where $r>d$; while, on the low-$r$ side of $d$ ($0<r<d$), we have a polynomial tail
given by the first derivative of a Legendre polynomial of degree $2p$ in $r/d$. 
This will change the background below the interatomic distances' peaks, due to the step-like contributions from the Heaviside functions. 
This is very evident in \fref{fig:pdf} for some example calculated {\gr}.

Legendre polynomials and their recursion are described in \sref{app2:LegeP}. 
The first few even Legendre polynomials with their first derivatives are listed here.
\[
\begin{array}{c|c|c}
p  & P_{2p}(x)  & \DSF{\mathrm{d}}{\mathrm{d}x} P_{2p}(x) \\
\ &&\\\hline
0  & 1  &  0 \\
 1 &  \DSF{1}{2}\lrb{3x^2-1} & 3x\\
 2 &   \DSF{1}{8}\lrb{35x^4-30x^2+3} &\DSF{5}{2}x\lrb{7x^2-3}\\
 3 &   \DSF{1}{16}\lrb{
 231x^6 -315x^4+105x^2-5
 } & 
 \DSF{21}{8} x \left(33 x^4-30
   x^2+5\right)
\\\ &\ & \\
\end{array}
\]

\section{Example calculations and graphics}\label{sec:GRAP}

\begin{figure}\centering
 \includegraphics[width=0.58\textwidth]{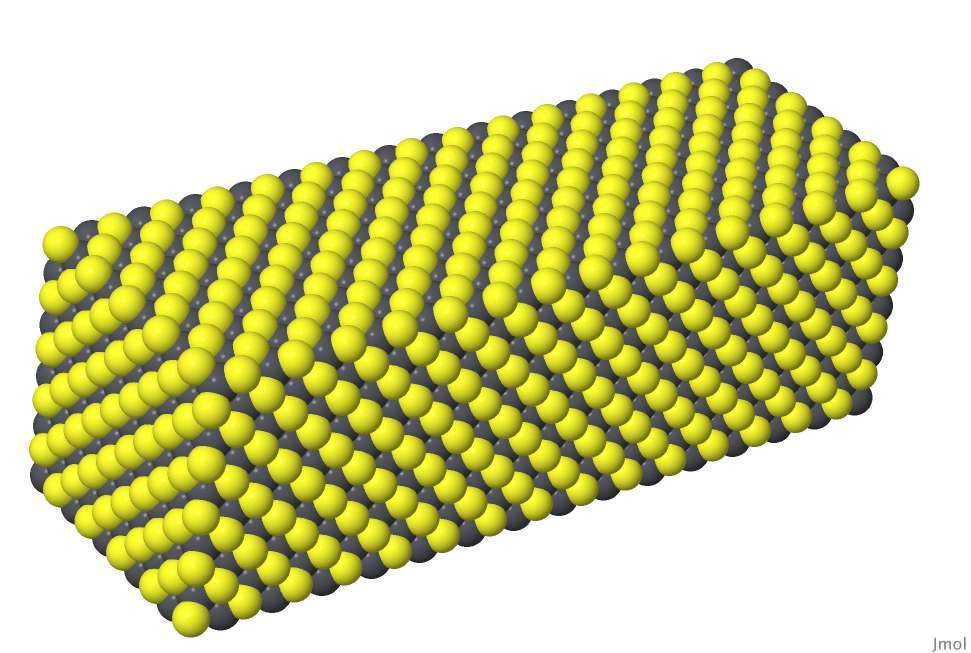}
\caption{The tetragonal prism of PbS (about $3\times 3\times 9$~nm) used as a model AO for graphical illustrations.}
\label{fig:AO1}
\end{figure}

To ease understanding of concepts here presented, we have made some example calculations, building first an 
ideal AO in the form of a NC of PbS (space group $Fm{\overline{3}m}$, lattice parameter 5.936~\AA, rock-salt structure), with the shape of a parallelogram of $5\times5\times 15$ unit cells (\fref{fig:AO1}). Special attention has been devoted to building the surface in a way that does not reduce the overall symmetry; however, the point group of the AO as a whole is tetragonal (due to the elongated shape), more precisely $D_{4h}$ or $4/mmm$. 

{\REVISED{We like to stress that this symmetry reduction has a deep meaning. In fact, it correlates the ODF with the particle shape. 
Now, given that the particle shape is often the main reason for an anisotropic ODF, this is surely an advantage. Even more so when there is also a correlation between shape and strain. }}

We will hereafter show patterns of PbS modified by ODF for the three common experimental geometries (DS, BB, FP) illustrated in \sref{samsym}. 
Looking at  \tref{tbl:1}, we can see that the selection rules for $4/mmm$ AO symmetry allow only coefficients $Z_{2p;m,0}$ where $m$ is a non-negative multiple of 4, obeying $m\leqslant l=2p$, to be nonzero. The standard DSE includes only the term $(l,m)=(0,0)$. 
With texture, higher orders up to $l=12$ are limited to $(l,m)=(2,0),\,(4,0),\ldots,(12,0)$; $(4,4),\,(6,4),\,\ldots,(12,4)$; $(8,8),\,(10,8),\,(12,8)$; and $(12,12)$. So, the dis-uniformity of the ODF is described by a grandtotal of 15 terms up to order $l=12$, that is fairly high. 

We shall here also simplify the treatment of atomic form factors. So, instead of the $q$-dependent expression $\RE\lrb{{\overline{f}}_jf_k}$ 
for the scattering product of the $(j,k)$-th pair of atoms, we will use the simpler form $Z_jZ_k$, the product of the atomic numbers (82 for Pb and 16 for S). We shall also set the scale by dividing each pattern by the self-scattering term 
\[
\mathop{\sum}_{j=1}^N\lrv{f_j}^2=\mathop{\sum}_{j=1}^NZ_j^2
\]
in order to set a common scale. That means, our plots will be all of the (modified) {\sq}. 

\begin{figure}\centering
 \includegraphics[width=0.9\textwidth]{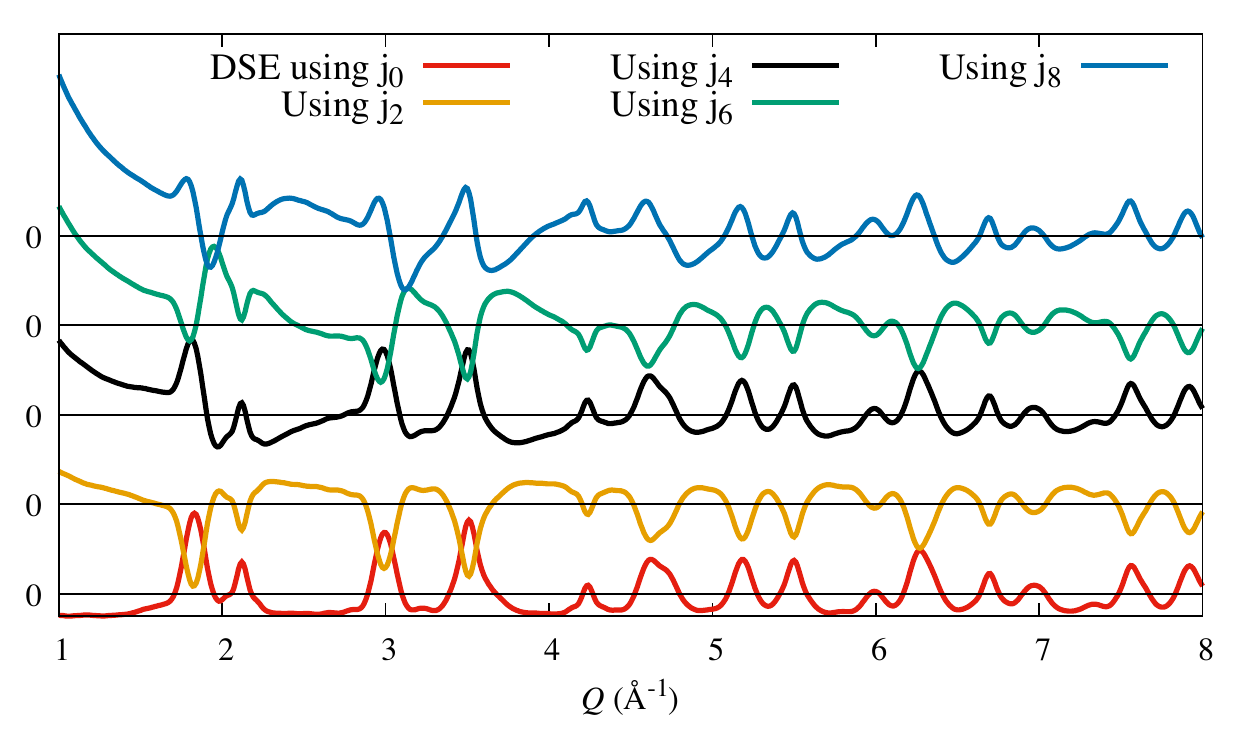}
\caption{The standard DSE-calculated {\sq} pattern (that for uniform ODF) and a few ``higher-order'' variants where we simply substitute $j_{l}(x)$ for $j_{0}(x)$. It is interesting the similarity -especially in the peaked regions - apart from the  alternating sign.}
\label{fig:SUPORD}
\end{figure}

Firstly, just to have an impression about the superior spherical Bessel function terms, as a play, 
we will compute the standard DSE substituting $j_{2p}(x)$ $(p=1,2,\ldots)$ for $j_{0}(x)=\sin(x)/x$. 
To be noted in \fref{fig:SUPORD} is the striking similarity between the modified DSE patterns at different orders, apart from a sign $(-1)^p$. 

Next, we make some true example calculations based on the same AO, using \eeeref{DS1}{BB1}{FP1}. 
As the values of the 15 allowed coefficients $Z_{l;m,0}$ ($l\leqslant 12$) are arbitrary within limits in \eref{maxvalZ}, we evaluated 
- for each of the three experimental geometries - 15 {\sq} patterns, each one modified by ``switching on'' a single  $Z_{l;m,0}$. 
Each time, we both add and subtract the chosen perturbing term, fixing the respective coefficient $Z_{l;m,0}=\pm \kappa(2l+1)$.  
We let $\kappa=1$ for the DS and FP geometry, where perturbations are weaker; we set it to 0.2 for the BB, in order to avoid huge negative intensity values. Of course, any good refinement program would determine coefficient values that reproduce the observed {\sq}, 
so this is not a problem in applications. 
\begin{figure}\centering
 \includegraphics[width=0.8\textwidth]{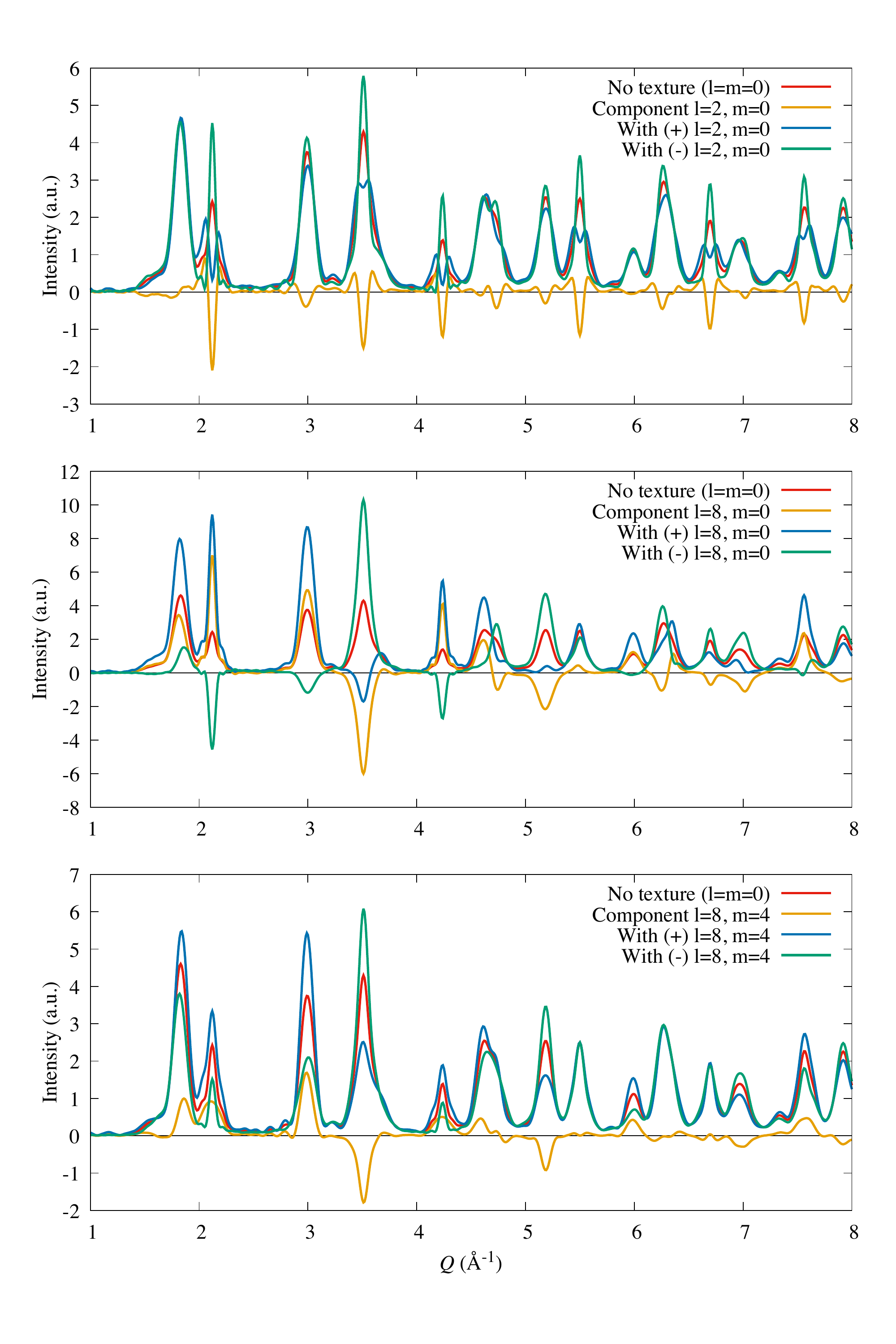}
\caption{Debye-Scherrer (capillary) geometry: red - {\sq} without texture, orange - the single texture component, blue - {\sq} with component added, green - {\sq} with component subtracted. The component magnitude was set as $Z_{l;m,0}=\pm (2l+1)$, its upper bound from \eref{maxvalZ}. This leads to negative intensity in some regions of plot (b) when the $(-)$ sign is used, so in reality bounds on $Z_{l;m,0}$ must be tighter.(top) {\sq} and its modification with the $(2,0)$ component; (middle) with the $(8,0)$ component; (bottom) with the $(8,4)$ component.}
\label{fig:DSpert}
\end{figure}
\begin{figure}\centering
 \includegraphics[width=0.8\textwidth]{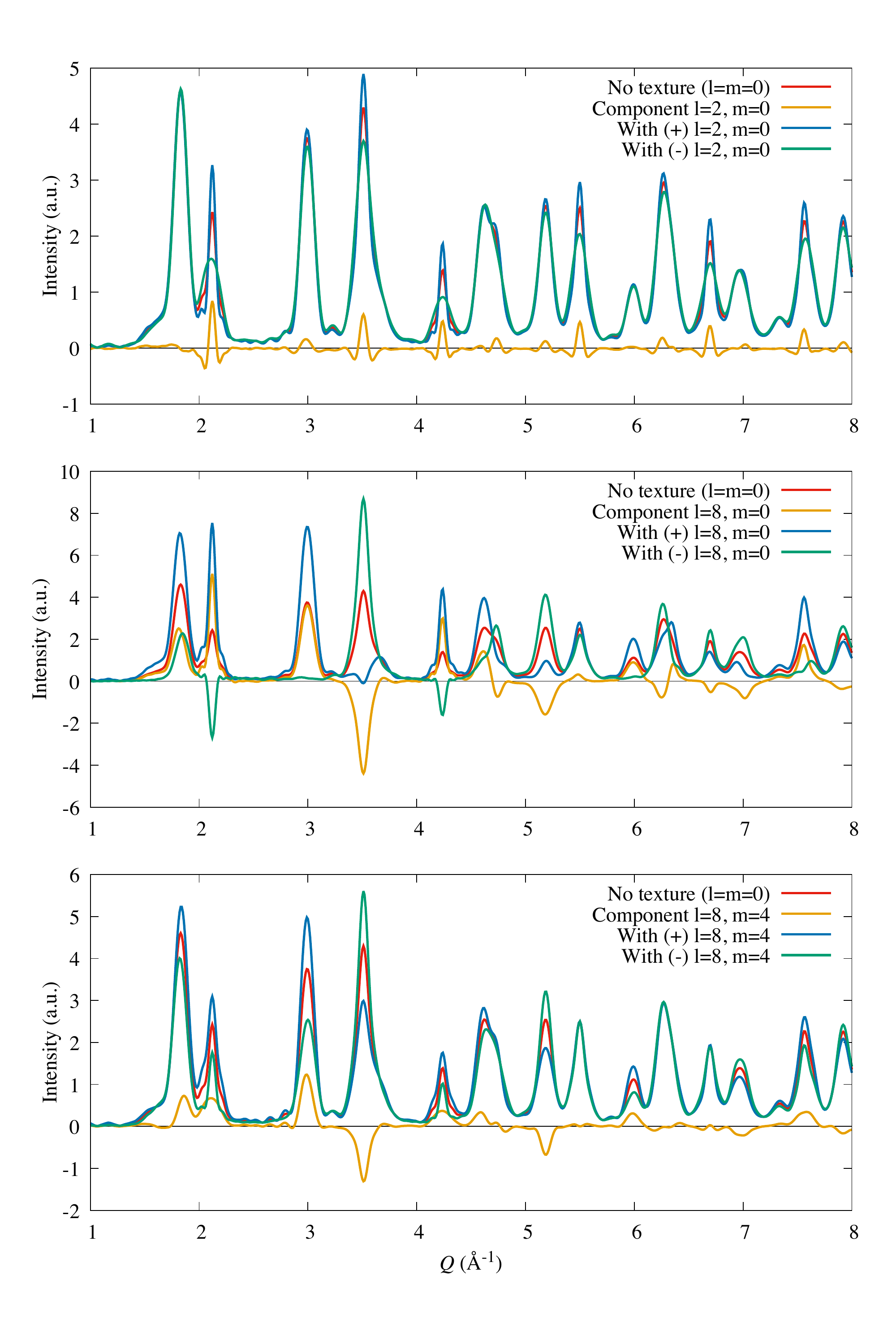}
\caption{Bragg-Brentano (reflection) geometry. Same color scheme as \fref{fig:DSpert}. 
The component magnitude was set as $Z_{l;m,0}=\pm 0.2 (2l+1)$, or 1/5 of its upper bound from \eref{maxvalZ}. Still we can see some negative intensity in some regions of plot (b) when the $(-)$ sign is used, so an even tighter bound is necessary in this case.
(top) {\sq} and its modification with the $(2,0)$ component; (middle) with the $(8,0)$ component; (bottom) with the $(8,4)$ component.}
\label{fig:BBpert}
\end{figure}
\begin{figure}\centering
 \includegraphics[width=0.8\textwidth]{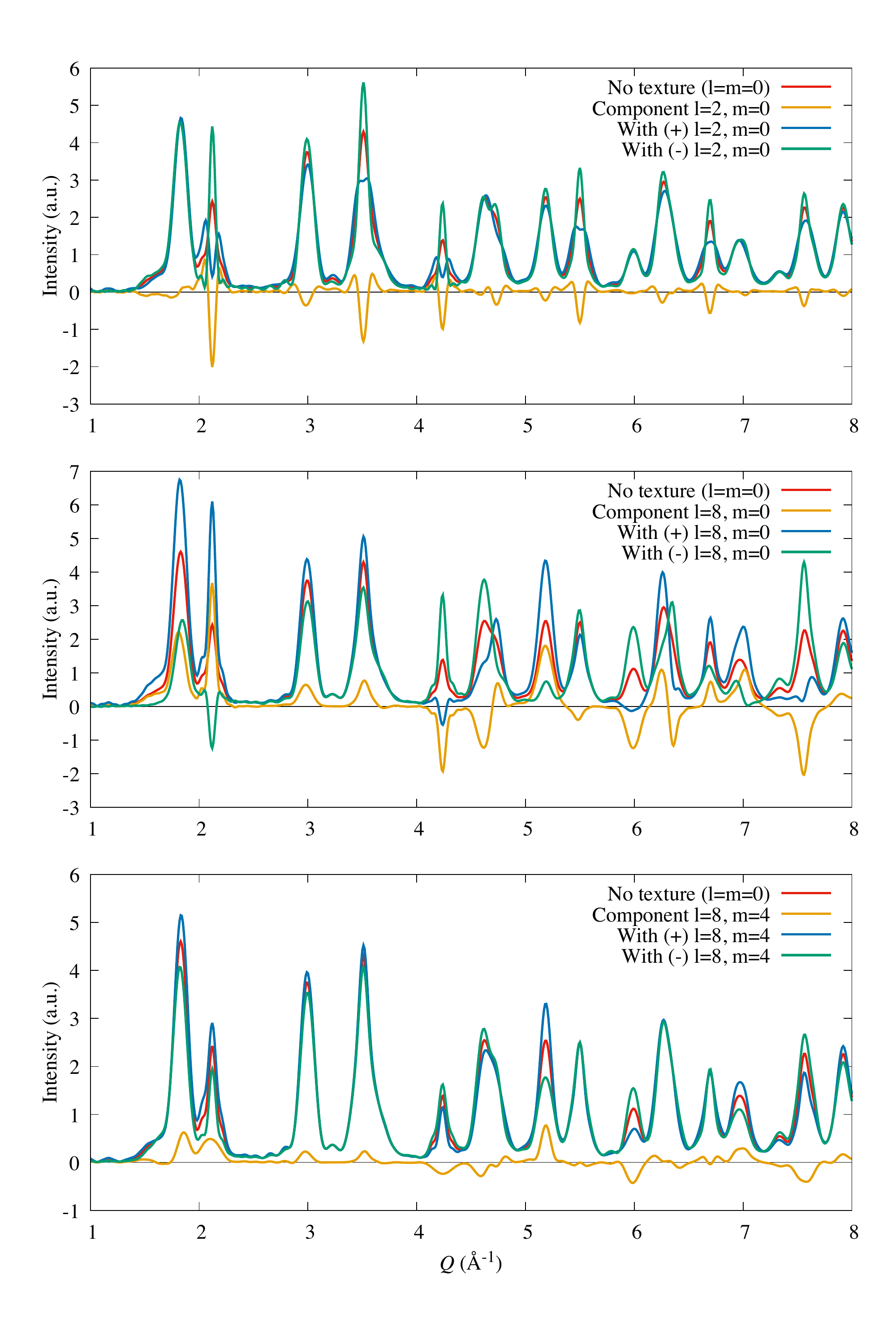}
\caption{Flat-plate (transmission) geometry. Same color scheme as \fref{fig:DSpert}. Here again we set $Z_{l;m,0}=\pm (2l+1)$, its upper bound from \eref{maxvalZ}. Negative intensity regions again indicate the need for tighter bounds on the coefficients. (Top) {\sq} and its modification with the $(2,0)$ component; (middle) with the $(8,0)$ component; (bottom) with the $(8,4)$ component.}
\label{fig:FPpert}
\end{figure}

Observing the graphs in \fref{fig:DSpert}, \fref{fig:BBpert}, \fref{fig:FPpert} we can note several interesting features. 
Perhaps the most important is that texture - at least when combined with some shape anisotropy, as in this example - 
does not just modify the intensity of Bragg peaks but changes their shape as well. Apparent peak splits can be seen in \fref{fig:DSpert} (a) 
(DS geometry, case $(l,m)=(2,0)$), for instance; other graphs show apparent peak shifts, broadening or narrowing and profile alterations. 
Furthermore, though relatively smaller, changes in the background can be observed as well. This point is important because shape anisotropy is 
very often accompanying - and likely causing - texture in powder samples. 
Therefore, we believe that it is important to have new tools as the modified DSE here presented, in order to account 
precisely for all effects of texture combined with size and shape anisotropy.

{\REVISED{
\subsection{Comparison with Bragg methods}\label{BMC}
A comparison with traditional Bragg (Rietveld) methods of calculating a powder diffraction pattern and SH components is shown in \fref{fig:TS-Riet}. To do so, we computed the texture-free pattern 
for one spherical NC with $D=5.4$~nm using the TOPAS software (\citeasnoun{TOPAS_Coelho_2018}) implementing the SH as in \citeasnoun{Jarvinen_1993}. This sphere has equivalent volume to the rod used so far. We selected the Debye-Scherrer geometry for this comparison.\\ 
TOPAS uses the Laue symmetry $m{\overline{3}}m\equiv O_h$ (as derived from the given {\texttt{.cif}} file). 
Maybe it would have been possible to force the tetragonal group onto TOPAS, but still the shape anisotropy would have been more difficult. Therefore we produces another simulation, using a cubis PbS NC with 
$7\times7\times7$ unit cells and approximately the same volume as above. We also assigned this time the 
$m{\overline{3}}m\equiv O_h$ group as AO symmetry. We calculated some of the corresponding Kubic harmonics ($K_4^1$, $K_6^1$, $K_8^1$) with both programs. The results are shown in \fref{fig:TS-Riet}. 
The patterns without texture (top quadrant) match very nicely. 
Minor differences are in the (low) diffuse background of the total scattering pattern. This however can be expected because a) the actual shape difference (cube \emph{vs.} sphere) must yield some small differences, b) what lies ``underneath the Bragg peaks'' is where total scattering and Rietveld-Bragg methods are supposed to differ.
The ODF contributions (we show only two) match very closely indeed, as it can be seen in the middle and bottom quadrant. 
%
\begin{figure}\centering
 \includegraphics[width=0.8\textwidth]{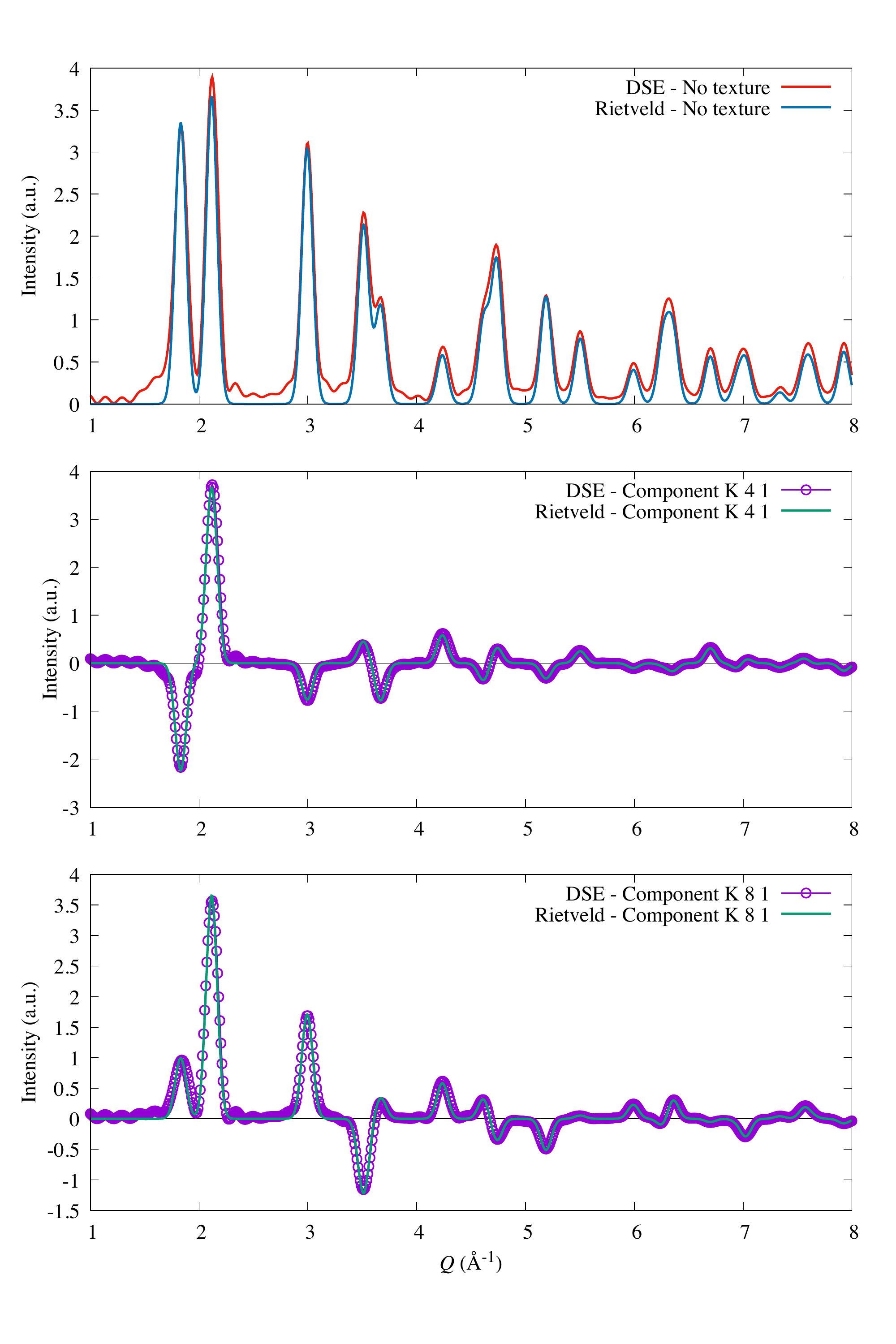}
\caption{Debye-Scherrer (capillary) geometry: top - {\iq} without texture for different shapes as calculated by DSE (red) and Rietveld (blue); middle - kubic $K_4^1$ component as calculated by DSE (violet, dotted) and Rietveld (green); bottom -kubic $K_8^1$ component as calculated by DSE (violet, dotted) and Rietveld (green).}
\label{fig:TS-Riet}
\end{figure}
}}

\subsection{Example calculations of {\gr} Pair Distribution Function}

\begin{figure}\centering
 \includegraphics[width=0.8\textwidth]{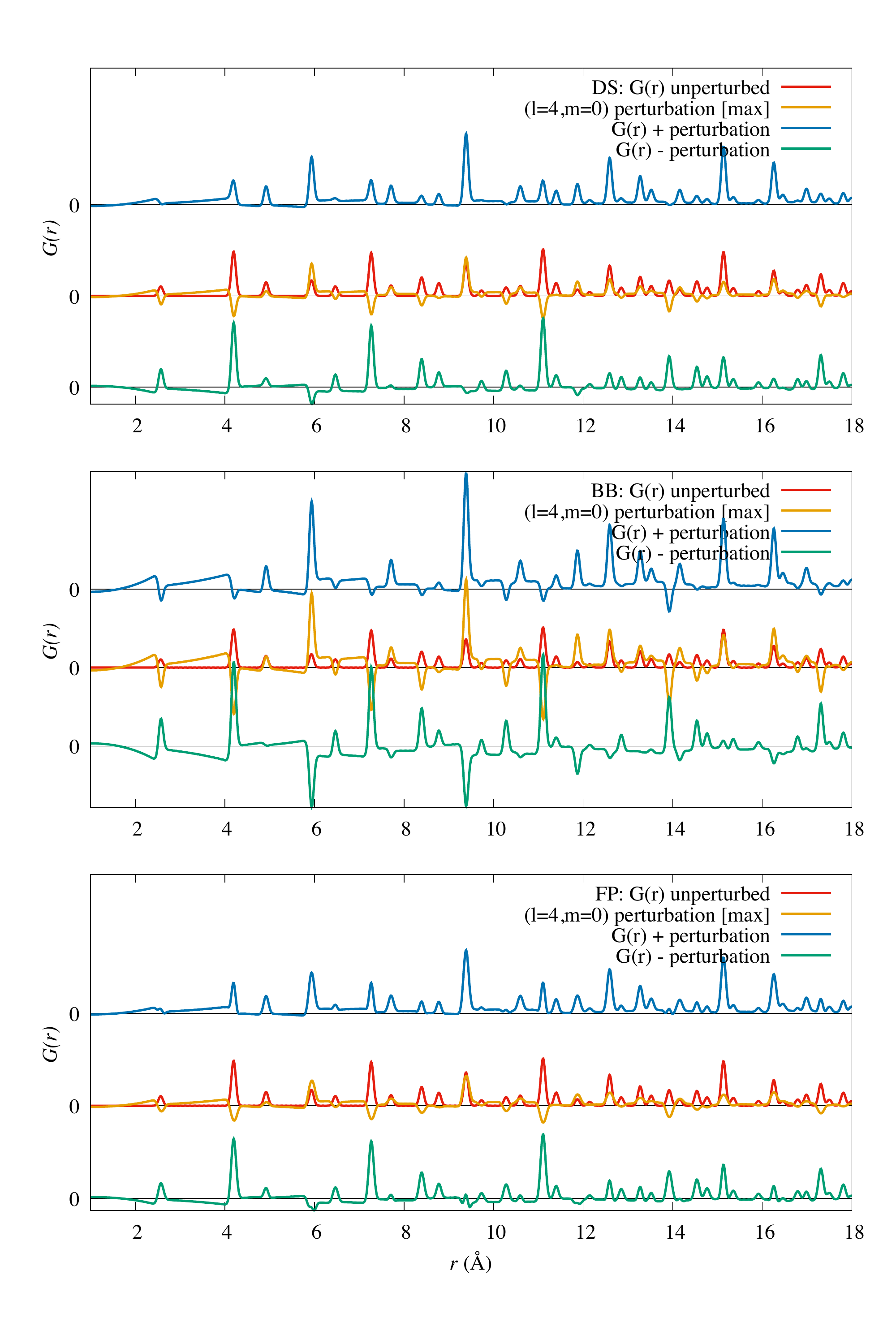}
\caption{The {\gr} unperturbed (red) and the same perturbed by $\pm$ the maximal $(4,0)$ perturbation (green for $(-)$ sign, blue for $(+)$). Plots are shifted vertically for convenience. The separate perturbation is plotted in orange over the unperturbed function. (Top) - for DS geometry; (middle) - for BB (note the much higher effect); (bottom) - for FP.}
\label{fig:pdf}
\end{figure}

Finally, in order to verify various points discussed in \sref{sec:pdf}, we show (\fref{fig:pdf}) the plot of one 
calculated {\gr} for the same PbS NC and for the 3 experimental geometries. The {\sq} was evaluated up to $Q=60$ (fairly high), and a Gaussian broadening of r.m.s. width 0.05~\AA~ 
was added to the interatomic distances, like a moderate Debye-Waller factor. This is very effective in regularising the {\gr}, as it is well known. 
Similar vibrational amplitudes are common in ordinary matter. 
Again, as before, we calculated the unperturbed {\sq} via the DSE, then added/subtracted texture $(l,m)$ perturbations one at a time, 
always with the maximum coefficient allowed $\pm(2l+1)$.
We chose the lowest order of texture whose effects were visible in the graph. It turned out that we did not have to go far, 
as at $(l,m)=(4,0)$ every possible effect proposed in \sref{sec:pdf} is already easily visible and likely making the analyisis quite complicated. Within the $\pm(2l+1)$ range of coefficients, several distance peaks may easily be deleted (in our plots become negative, therefore with lower coefficient magnitude they must go to 0). Furthermore, the background is rich of ramps and steps due to the 
expansion of superior spherical Bessel functions in the $j_0(Qr)$ basis, as explained in \sref{sec:pdf}. \\
Again, from \fref{fig:pdf}, we can very well see that the BB geometry is much more affected by texture with respect to the other systems. 
This may be an useful tip when planning experiments on samples under suspicion of preferred orientation.
%

\section{Conclusions}
We have derived extended Debye scattering equations that encode sums over higher-order even spherical Bessel functions which account for  corrections to moderate texture.  We showed that, as in the Bragg scattering case, the texture effect modifies the diffraction maxima intensities, possibly leading to their cancellation. We extended our approach to the direct space transforms deriving one expression for the {\gr} function,  showing two important facts: {\emph i}) the well-known texture effect in reciprocal space has its dual counterpart in real space where the height of many interatomic distance peaks will change, in particular  for some combination of texture coefficients some distance peaks may disappear; and  {\emph{ii}}) the contribution from the higher-order even spherical Bessel functions will introduce also polynomial tails on the low-$r$ side of the peaks, with a step to zero at $r\geqslant d$ which could be difficult to model.

\clearpage\newpage

\appendix
\section{Texture special functions: definitions and notation}\label{app}

\subsection{Legendre polynomials}\label{app2:LegeP}
Legendre polynomials definition:
\begin{equation}
P_l(x)=\DSF{(-1)^l}{2^ll!}\DSF{\partial^{l}}{\partial x^{l}}\lrb{1-x^2}^l
\end{equation}
Orthogonality:
\begin{equation}
\int_{-1}^1\DD{x}\ P_l(x)P_{l'}(x)=\int_{0}^{\pi}\sin\lrb{\phi}\DD{\phi}
\ P_l(\cos\lrb{\phi})P_{l'}(\cos\lrb{\phi})=\DSF{2}{2l+1}\delta_{l,l'}
\end{equation}
Recurrence:
\begin{equation}
P_{l+1}(x)= \DSF{2l+1}{l+1}xP_l(x)-\DSF{l}{l+1}P_{l-1}(x)
\end{equation}

\subsection{Associated Legendre functions}\label{app3:LegeF}
Associated Legendre functions: there are two definitions in literature.
\begin{eqnarray}
\text{(I):}&\ &\DST{ {}^{(\mathrm{I})\!\!}P_l^m\lrb{x}=\DSF{\lrb{-1}^{m+l}}{2^ll!}\lrb{1-x^2}^{m/2}\DSF{\partial^{l+m}}{\partial x^{l+m}}\lrb{1-x^2}^l;}\\
\text{(II):}&\ &\DST{ {}^{(\mathrm{II})\!\!}P_l^m\lrb{x}=\DSF{\lrb{-1}^{l}}{2^ll!}\lrb{1-x^2}^{m/2}\DSF{\partial^{l+m}}{\partial x^{l+m}}\lrb{1-x^2}^l=\lrb{-1}^m{}^{(\mathrm{I})\!\!}P_l^m\lrb{x}
.}
\end{eqnarray}
Definition (II) is in \citeasnoun{Edmonds}, \citeasnoun{Messiah1}, \citeasnoun{MasterRichards98}, \citeasnoun{NikifUva} - most used in Quantum Mechanics. 
Here we shall adopt (I), while using Ref.~\citeasnoun{AbraStegLegendre}'s convention:
\begin{eqnarray}
P_l^m\lrb{x}&\equiv& {}^{(\mathrm{I})\!\!}P_l^m\lrb{x};\nonumber\\
P_{lm}\lrb{x}&\equiv& {}^{(\mathrm{II})\!\!}P_l^m\lrb{x}=(-1)^mP_l^m\lrb{x}.
\end{eqnarray}

In both cases, 
\begin{equation}
P_l^{-m}\lrb{x}=(-1)^m\DSF{\lrb{l-m}!}{\lrb{l+m}!}P_l^m\lrb{x}
\end{equation}
\begin{equation}
P_l^{0}\lrb{x}=P_l\lrb{x}, \qquad\text{where}\ P_l\lrb{x}=
\DSF{\lrb{-1}^{l}}{2^ll!}\DSF{\partial^{l}}{\partial x^{l}}\lrb{1-x^2}^l
\end{equation}
Orthogonality:
\begin{equation}
\int_{-1}^1\DD{x}\ P_l\lrb{x}P_{l'}\lrb{x} = \DSF{2}{2l+1} \ \delta_{l,l'}
\end{equation}
\begin{equation}
\int_{-1}^1\DD{x}\ P_l^m\lrb{x}P_{l'}^m\lrb{x} = \DSF{2}{2l+1} \DSF{\lrb{l+m}!}{\lrb{l-m}!}\ \delta_{l,l'}
\end{equation}
Recurrence over $l$:
\begin{equation}
P_l^m(x)=\DSF{2l-1}{l-m}xP_{l-1}^m(x)-\DSF{l+m-1}{l-m}P_{l-2}^m(x)\qquad\text{when}\ m<l;
\end{equation}
for $m=l$, we complete with 
\[
P_l^l(x)=(-1)^l(2l-1)!!(1-x^2)^{l/2}\qquad \left|\qquad (2l-1)!!\equiv\mathop{\prod}_{j=1}^l(2j-1)\right.
\]

\subsection{Spherical harmonics}\label{app4:SPH}
Spherical harmonics can be defined in two ways, depending on the choice for Legendre functions:
\begin{eqnarray}
\text{(I):}&\ &\DST{Y_l^m\lrb{\theta,\phi}=\sqrt{\DSF{2l+1}{4\pi}}
\lrb{\DSF{\lrb{l-m}!}{\lrb{l+m}!}}^{1/2}P_l^m\lrb{\cos\lrb{\theta}}
\EE^{\IMA m\phi}
}\\
\text{(II):}&\ &\DST{Y_l^m\lrb{\theta,\phi}=\lrb{-1}^m\sqrt{\DSF{2l+1}{4\pi}}
\lrb{\DSF{\lrb{l-m}!}{\lrb{l+m}!}}^{1/2}P_{lm}\lrb{\cos\lrb{\theta}}
\EE^{\IMA m\phi}
}
\end{eqnarray}
with the $\lrb{-1}^m$ sign appearing in (II) is the Condon-Shortley phase, compensating for the 
absence of a similar factor in the definition of $P_{lm}$, as discussed above. 
The SH are orthonormal with respect to both indexes:
\begin{equation}
\int_{0}^{2\pi}\DD{\phi}\int_{0}^{\pi}\sin\lrb{\theta}\DD{\theta}\ 
Y_l^m\lrb{\theta,\phi}
{\overline{Y}}_{l'}^{m'}\lrb{\theta,\phi}=\delta_{l,l'}\delta_{m,m'}
\end{equation}

Note that Bunge \citeasnoun{Bunge82} defines the SH slightly differently. He uses the symbol $\kappa_l^m\lrb{\Phi,\beta}$ instead of 
$Y_l^m\lrb{\theta,\phi}$ ($\theta\equiv\Phi,\phi\equiv\beta$) and
\[
\kappa_l^m\lrb{\Phi,\beta}=\EE^{\IMA m \beta}\DSF{\lrb{-1}^{l+m}}{2^ll!}\sqrt{\DSF{2l+1}{4\pi}}\lrb{\DSF{\lrb{l+m}!}{\lrb{l-m}!}}^{1/2}
\lrb{1-x^2}^{-m/2}
\DSF{\partial^{l-m}}{\partial x^{l-m}}\lrb{1-x^2}^l\Bigl.\Bigr|_{x=\cos\lrb{\Phi}}
\]
from Eqs.~(14.38),~(14.39) of \citeasnoun{Bunge82}.
We can use the identity \citeasnoun{NikifUva}, 
\[
\DSF{\partial^{l-m}}{\partial x^{l-m}}\lrb{1-x^2}^l
=
\lrb{-1}^m\lrb{1-x^2}^m
 \DSF{\lrb{l-m}!}{\lrb{l+m}!}\ 
\DSF{\partial^{l+m}}{\partial x^{l+m}}\lrb{1-x^2}^l
\]
so we can write
\[
\kappa_l^m\lrb{\Phi,\beta}=\EE^{\IMA m \beta}\DSF{\lrb{-1}^{l}}{2^ll!}\sqrt{\DSF{2l+1}{4\pi}}\lrb{\DSF{\lrb{l-m}!}{\lrb{l+m}!}}^{1/2}
\lrb{1-x^2}^{m/2}
\DSF{\partial^{l+m}}{\partial x^{l+m}}\lrb{1-x^2}^l
\Bigl.\Bigr|_{x=\cos\lrb{\Phi}}
\]
which leads to
\[
\kappa_l^m\lrb{\Phi,\beta}=\lrb{-1}^mY_l^m\lrb{\Phi,\beta}={\overline{Y}_l^{-m}}\lrb{\Phi,\beta}
\]

\section{Computational wisdom}\label{appB}

\subsection{General double-step recurrence}\label{app:step2}

{\REVISED{Many functions and polynomials useful in the computation of spherical harmonics 
are defined recursively. In crystallography, we often have to consider only the even terms of such recurrence. 
As it is clearly a waste to compute the odd terms when we are interested only in the even ones, 
we give here a recipe to transform a three-term recurrence relation into one that uses only even orders. 
If we have a three-term recurrence relation}}
\[
V_{m+1}=a_mV_m+b_mV_{m-1}
\]
we can expand $V_m$ in the RHS:
\[
V_{m+1}=a_m\lrb{a_{m-1}V_{m-1}+b_{m-1}V_{m-2}}+b_mV_{m-1}=\lrb{a_ma_{m-1}+b_m}V_{m-1}+a_mb_{m-1}V_{m-2}
\]
then, considering also
\[
a_{m-2}V_{m-2}=V_{m-1}-b_{m-2}V_{m-3}
\]
and substituting,
\begin{eqnarray}
V_{m+1}&=&\lrb{a_ma_{m-1}+b_m}V_{m-1}+a_mb_{m-1}\DSF{V_{m-1}-b_{m-2}V_{m-3}}{a_{m-2}}\nonumber\\
&=&\lrb{a_ma_{m-1}+b_m+\DSF{a_mb_{m-1}}{a_{m-2}}}V_{m-1}-\DSF{a_mb_{m-1}b_{m-2}}{a_{m-2}}\nonumber\\
&=&\DSF{a_m\lrb{a_{m-1}a_{m-2}+b_{m-1}}+b_ma_{m-2}}
{a_{m-2}}
V_{m-1}-\DSF{a_mb_{m-1}b_{m-2}}{a_{m-2}}
V_{m-3}
\end{eqnarray}

\subsection{Clenshaw recurrence}
\label{app:Clen}

Suppose we need to evaluate linear combinations of the form 
\begin{equation}
f(x)=\mathop{\sum}_{k=0}^N\,c_kF_k(x)\label{clin}
\end{equation}
where the $F_k(x)$ obey a three-term recurrence
\begin{equation}
F_{n+1}(x)=\alpha_n(x)F_n(x)+\beta_n(x)F_{n-1}(x)
\end{equation}
and the first two terms $F_0,F_1$ are known 
[note that a special case of \eref{clin} is the evaluation of  $F_N(x)$, just setting $c_0=\ldots=c_{N-1}=0$ and $c_N=1$].

The most efficient way to compute such linear combinations is usually Clenshaw's recurrence (\citeasnoun{Clenshaw62}, \citeasnoun{NRClenshaw}), using auxiliary functions $y_k(x)$. In simple terms, we set
\begin{eqnarray}
y_{N+2}(x)&\equiv&0;\qquad
y_{N+1}(x)\equiv 0;\\
y_k(x)&=&\alpha_k(x)y_{k+1}(x)+\beta_{k+1}(x)y_{k+2}(x)+c_k,\qquad k=N,N-1,\ldots, 1
\nonumber
\end{eqnarray}
and at the end it can be shown that
\begin{equation}
f(x)=\lrb{c_0+y_2(x)\beta_1(x)}F_0(x)+y_1(x)F_1(x)
\end{equation}
This is precise and does not require evaluating the $F_k(x)$ first, as only the first two are used. 

In the rare cases when
\begin{equation}
\lrv{f(x)}<\!\!<\lrv{\lrb{c_0+y_2(x)\beta_1(x)}F_0(x)}+\lrv{y_1(x)F_1(x)}
\end{equation}
then the opposite procedure is better:
\begin{eqnarray}
y_{-2}(x)&\equiv&0;\nonumber\\
y_{-1}(x)&\equiv&0;\\
y_k(x)&=&\DSF{-c_k-\alpha_k(x)y_{k-1}+y_{k-2}}{\beta_{k+1}(x)},\qquad k=0,1,\ldots, N-1
\nonumber
\end{eqnarray}
and
\begin{equation}
f(x)=\lrb{c_N-y_{N-2}(x)}F_N(x)-\beta_N(x)y_{N-1}(x)F_{N-1}(x)
\end{equation}
Here of course the higher terms $F_N(x)$ and $F_{N-1}(x)$ have to be evaluated too.

%
%

\subsection{Recursive evaluation of spherical harmonics}

Here we follow \citeasnoun{MasterRichards98}. 
{\REVISED{The computation of SH of higher orders may be afflicted by numerical instabilities if 
direct formula computation is attempted. Moreover, while $l$-recursions based recursive methods may be 
trivial, the accompanying $m$-based recursion is not and may induce large numerical errors. 
Hereafter is the most precise and fastest way to manage such recursions. }}

Write
\[
Y_l^m(\theta,\phi)=\lrb{\sin\lrb{\theta}}^mW_l^m(\theta)\EE^{\IMA m\phi}
\]
We know that
\[
W_l^{l}(\theta)=(-1)^{l}\sqrt{\DSF{2l+1}{4\pi}}\sqrt{\DSF{1}{\lrb{2l}!}}\lrb{2l-1}!!
\]
where
$
\lrb{2l-1}!!\equiv 2^{-l}\,{\lrb{2l}!}/l!
$
and
\[
W_l^{l-1}(\theta)=
(-1)^{l-1}
\sqrt{\DSF{2l+1}{4\pi}}
\sqrt{\DSF{1}{\lrb{2l-1}!}} \lrb{2l-1}!!\cos\lrb{\theta}
=-\sqrt{2l}\ \cos\lrb{\theta}   
\ 
W_l^{l}(\theta)
\]
{\REVISED{Now we can give a stable downward $m$-recurrence for $m$ from $l$ to 0, 
compatible with Clenshaw's method of \sref{app:Clen}, as
\[
W_l^{m-1}(\theta)=-\ \DSF{2m\cos\lrb{\theta}}{\sqrt{\lrb{l+m}\lrb{l-m+1}}}\ W_l^m(\theta)\ 
-\ \DSF{\sqrt{\lrb{l-m}\lrb{l+m+1}}\ \sin^2\lrb{\theta} }{\sqrt{\lrb{l+m}\lrb{l-m+1}}}\ W_l^{m+1}(\theta)
\]
The SH (also with negative $m$ values) are then obtained as
\[
Y_l^m(\theta,\phi)=\sin^m\lrb{\theta}\EE^{\IMA m\phi}W_l^m(\theta);
\qquad
Y_l^{-m}(\theta,\phi)=(-1)^m\sin^m\lrb{\theta}\EE^{-\IMA m\phi}W_l^m(\theta);
\]
}}
\clearpage

\end{document}